\documentclass[aip,twocolumn,10pt]{revtex4}%
\usepackage{amsmath}
\usepackage{amsfonts}
\usepackage{amssymb}
\usepackage{graphics}
\usepackage{graphicx}
\providecommand{\U}[1]{\protect\rule{.1in}{.1in}}
\begin{document}
\title{Active protection of a superconducting qubit with an interferometric Josephson isolator}
\author{Baleegh Abdo}
\author{Nicholas T. Bronn}
\author{Oblesh Jinka}
\author{Salvatore Olivadese}
\author{Antonio D. C\'{o}rcoles}
\author{Vivekananda P. Adiga}
\author{Markus Brink}
\author{Jerry M. Chow}
\address{IBM T. J. Watson Research Center, Yorktown Heights, New York 10598, USA}
\author{Russell E. Lake}
\author{Xian Wu}
\author{David P. Pappas}
\affiliation{National Institute of Standards and Technology, Boulder, Colorado 80305, USA}
\date{\today}

\begin{abstract}	
 Nonreciprocal microwave devices play several critical roles in high-fidelity, quantum-nondemolition (QND) measurement schemes. They separate input from output, impose unidirectional routing of readout signals, and protect the quantum systems from unwanted noise originated by the output chain. However, state-of-the-art, cryogenic circulators and isolators are disadvantageous in scalable superconducting quantum processors because they use magnetic materials and strong magnetic fields. Here, we realize an active isolator formed by coupling two nondegenerate Josephson mixers in an interferometric scheme. Nonreciprocity is generated by applying a phase gradient between the same-frequency pumps feeding the Josephson mixers, which play the role of the magnetic field in a Faraday medium. To demonstrate the applicability of this Josephson-based isolator for quantum measurements, we incorporate it into the output line of a superconducting qubit, coupled to a fast resonator and a Purcell filter. We also utilize a wideband, superconducting directional coupler for coupling the readout signals into and out of the qubit-resonator system and a quantum-limited Josephson amplifier for boosting the readout fidelity. By using this novel quantum setup, we demonstrate fast, high-fidelity, QND measurements of the qubit while providing more than $20$ dB of protection against amplified noise reflected off the Josephson amplifier.  
\end{abstract}

\maketitle

\newpage
The capability to perform fast, high-fidelity, single-shot, quantum nondemolotion (QND) measurement of qubits is a critical requirement for the operation of quantum computers \cite{DevoretScience}. It is needed, for instance, to read out qubits in real-time \cite{QuantumJumps,QubitJPC}, track the evolution of quantum states \cite{stabilizetrajectory,MurchSingleTrajectory}, detect error syndromes \cite{SunTrackPhotonJumps}, stabilize quantum states \cite{VijayNature,feedbackStabilization}, and apply quantum feedback, as required in certain protocols \cite{FeedbackJPC,JPADicarloReset,ContQuantErrCorr}. In the case of superconducting quantum processors, one prominent platform for performing QND measurements is circuit quantum electrodynamics (cQED) \cite{cQEDBlais}, in which superconducting qubits are dispersively coupled to superconducting microwave readout resonators, and the qubit state is inferred by measuring the phase shift experienced by a weak near-resonance microwave signal applied to the readout resonator \cite{StrongCouplingCPB}. To perform such fast QND measurements in the cQED architecture, several key microwave components are commonly employed \cite{RapidQND}, such as: (1) quantum-limited Josephson amplifiers, which enhance the signal to noise ratio of the output chain while adding minimal amount of noise to the processed signals \cite{Caves,QuantumNoiseIntro}; (2) Purcell filters, which enable qubits to be coupled to fast readout resonators while inhibiting spontaneous emission of their excitations through the resonators \cite{Purcell,Bronn2015b}; and (3) nonreciprocal devices, which separate input from output and protect qubits against noise coming from the output chain \cite{QuantumJumps,QubitJPC}. However, unlike Josephson amplifiers or Purcell filters, which are compatible with superconducting circuits, have little or no internal loss, and can be miniaturized, nonreciprocal devices, i.e., cryogenic circulators and isolators, which are widely used in state-of-the-art quantum circuits, lack these desired properties. This is primarily because they rely on magneto-optical effects to break the transmission coefficient symmetry for light under exchanging sources and detectors, which entail using magnetic materials and strong magnetic fields \cite{Collin,Pozar}.      

In response to the immense challenge of eliminating these magnetic-based nonreciprocal devices, which severely hinder the scalability of superconducting quantum processors, a wide variety of viable alternative nonreciprocal schemes have been proposed and realized recently \cite{NoiselessCirc,KamalMetalmann,NonreciprocalResEng,JTWPA,TWPAthreewavemix,KIT,ReconfJJCircAmpl,NRAumentado1,NRAumentado2,circulatorDiVincenzo1,circulatorDiVincenzo2,Hallcirc,JDAQST,JDA,DircJPC,NonrecipMwOptoMech,MechOnChipCirc,DAMech,FreqConvIso,KamalSQUID,HFslug,circulatorLehnert}. Examples of these schemes include: Josephson traveling-wave parametric amplifiers \cite{JTWPA,TWPAthreewavemix}, kinetic-inductance traveling-wave parametric amplifiers \cite{KIT}, reconfigurable directional-amplifiers and circulators based on three-mode Josephson devices \cite{ReconfJJCircAmpl,NRAumentado1,NRAumentado2}, Hall-effect-based circulators \cite{circulatorDiVincenzo1,circulatorDiVincenzo2,Hallcirc}, interferometric Josephson directional amplifiers \cite{JDAQST,JDA,DircJPC}, circulators and directional amplifiers that are based on nanomechanical systems \cite{NonrecipMwOptoMech,MechOnChipCirc,DAMech}, Josephson-array transmission-line isolator \cite{FreqConvIso}, SQUID-based directional amplifiers \cite{KamalSQUID} or SQUID-variant devices, such as Superconducting Low-inductance Undulatory Galvanometer (SLUG) amplifiers \cite{HFslug}, and circulators which rely on synthetic-rotations generated using variable inductors \cite{circulatorLehnert}.    

Here, we introduce a new Josephson-based isolator, which is devoid of magnetic materials and strong magnets. The new device is fully compatible with superconducting circuits and can be integrated on chip. It breaks reciprocity by generating artificial gauge-invariant potential for microwave photons by parametrically modulating the inductive coupling between two modes of the system and achieves unidirectional transmission of propagating signals by creating constructive and destructive wave-interference between different paths in the device. The new isolator has two key differences, compared to other Josephson-based, microwave-controlled circulators realized recently \cite{ReconfJJCircAmpl,NRAumentado2}, it preserves the frequency of the routed quantum signals and can be operated by a single monochromatic microwave drive instead of three. These differences can result in a significant reduction in the overall control hardware resource for operating a larger number of devices.

Furthermore, here we take the additional crucial step of putting the new isolation scheme to the test in a quantum setup. More specifically, we couple it to a superconducting qubit in a fast, high-fidelity measurement setup and demonstrate that indeed it provides active protection to the qubit against unwanted noise coming from the output chain. 

\noindent\textbf{Results}

\begin{figure*}
	[tb]
	\begin{center}
		\includegraphics[
		width=1.7\columnwidth 
		]%
		{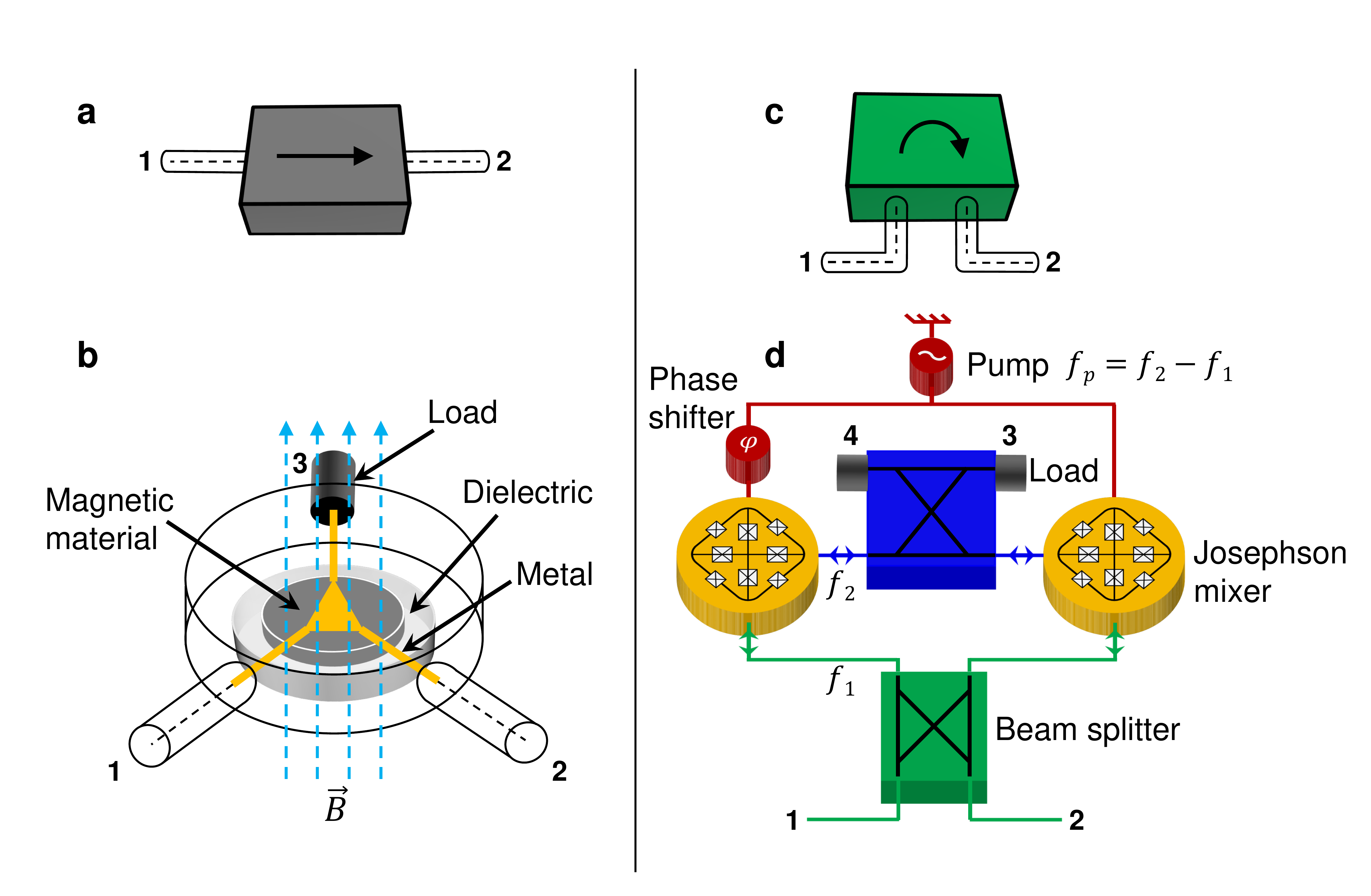}
		\caption{A microwave isolator, whose circuit symbol is shown in \textbf{a}, is a unidirectional, two-port microwave device. It transmits signals from port $1$ to $2$ (as indicated by the arrow) and blocks signals propagating in the opposite direction. To break the transmission-coefficient symmetry between source (input) and detector (output), commercial isolators rely on magneto-optical effects, which require magnetic materials and strong magnets. One common realization of such isolators is a three-port circulator in which one of its ports (i.e., $3$) is terminated by a $50$ $\Omega$ load as shown in \textbf{b}. In this example, signals entering port $1$ are directed towards port $2$, whereas signals entering port $2$ are dissipated in the matched load attached to the internal port $3$. \textbf{c} A circuit symbol for the Josephson-based isolator realized and measured in this work, which is based on a different reciprocity-breaking mechanism. In this device, schematically shown in \textbf{d}, two active Josephson mixers are coupled via beam-splitters and driven by a microwave pump source. Unidirectional transmission from port $1$ to $2$ is generated by constructive and destructive wave interference between different paths in the device. The interference pattern is controlled by the phase difference $\varphi=-\pi/2$ between the pumps feeding the two mixing stages. Signals propagating from port $2$ to $1$ are dissipated in the matched loads connected to the internal ports $3$ and $4$.          
		}
		\label{IsolatorJISComp}
	\end{center}
\end{figure*}

\noindent \textbf{Nonreciprocity mechanism.} To elucidate the basic idea behind the reciprocity-breaking mechanism of the new isolator, we qualitatively compare in Fig.\,\ref{IsolatorJISComp} a state-of-the-art magnetic-based isolator, whose circuit symbol is shown in Fig.\,\ref{IsolatorJISComp}a and the Multi-Path Interferometric Josephson ISolator (MPIJIS) realized and measured in this work, whose circuit symbol is introduced in Fig.\,\ref{IsolatorJISComp}c. While the circuit of the magnetic-based isolator can be realized in several different ways \cite{Pozar}, Fig.\,\ref{IsolatorJISComp}b exhibits a widely-used commercial realization, which captures the main common attributes of the device. In this realization, a two-port isolator is formed by terminating one port (e.g., port $3$) of a three-port, magnetic-based circulator with a matched load (e.g., $50$ $\Omega$). As seen in Fig.\,\ref{IsolatorJISComp}b, the circulator consists of a microstrip metallic junction connected to the center-conductor of the equally-spaced device ports and laid down on a disk-shaped dielectric substrate, which incorporates at its center a smaller ferrite disk functioning as a resonant cavity. The circulator circuit also includes a permanent magnet, not shown in the illustration, which induces a magnetic field in the device. As a result of the magnetic bias, the lowest-order resonant mode $f_0$ of the ferrite cavity splits into two modes having different frequencies $f_{\pm}$ and nonreciprocal azimuthal phase dependence $e^{\pm j\phi}$, which is set by the direction of the applied magnetic field. By engineering the amplitude of these modes via the bias field, a superposition pattern can be generated at the circulator ports, such that microwave signals at $f_1$, falling between the split-resonance frequencies $f_{-}<f_1<f_{+}$, propagate towards port $2$ upon entering port $1$, whereas signals entering $2$ propagate towards the internal port $3$ and dissipate their energy at the matched load \cite{Pozar}. 

The MPIJIS relies, in contrast, on different physics. In Fig.\,\ref{IsolatorJISComp}d, we depict the main components of the device. Unlike the passive magnetic-based isolator shown in Fig.\,\ref{IsolatorJISComp}b, the MPIJIS consists of two active Josephson mixers coupled together in an interferometric setup formed by two beam-splitters (i.e., $90^{\circ}$ hybrids). The two Josephson mixers function as lossless frequency converters (without photon gain) between two signals having frequencies $f_1$ (green) and $f_2$ (blue), which are supported by the resonance modes of the mixers. By construction, signals at $f_2$ only excite an internal mode of the system. To enable the frequency conversion, the Josephson mixers are driven by a monochromatic microwave pump at frequency $f_p=f_2-f_1$. Due to the parametric modulation of the Josephson mixers, an artificial gauge-invariant potential for microwave photons is generated, which depends on the pump phase difference $\varphi$ \cite{AhranovBohmMixers,AhranovBohmPhotonic,ReconfJJCircAmpl,NRAumentado2}. The induced potential consequently introduces a nonreciprocal phase shift for the frequency-converted signals \cite{JPCgyrator}, which is exploited in combination with wave interference (enabled by the beam-splitters) for transmitting signals in one direction, i.e., from port $1$ to $2$, and canceling signals in the opposite direction. Similar to the magnetic-based isolator (Fig.\,\ref{IsolatorJISComp}b), the canceled signals on the device port $1$ are directed towards the MPIJIS internal ports $3$ and $4$ where they are absorbed by the matched loads.

\begin{figure*}
	[tb]
	\begin{center}
		\includegraphics[
		width=1.7\columnwidth 
		]%
		{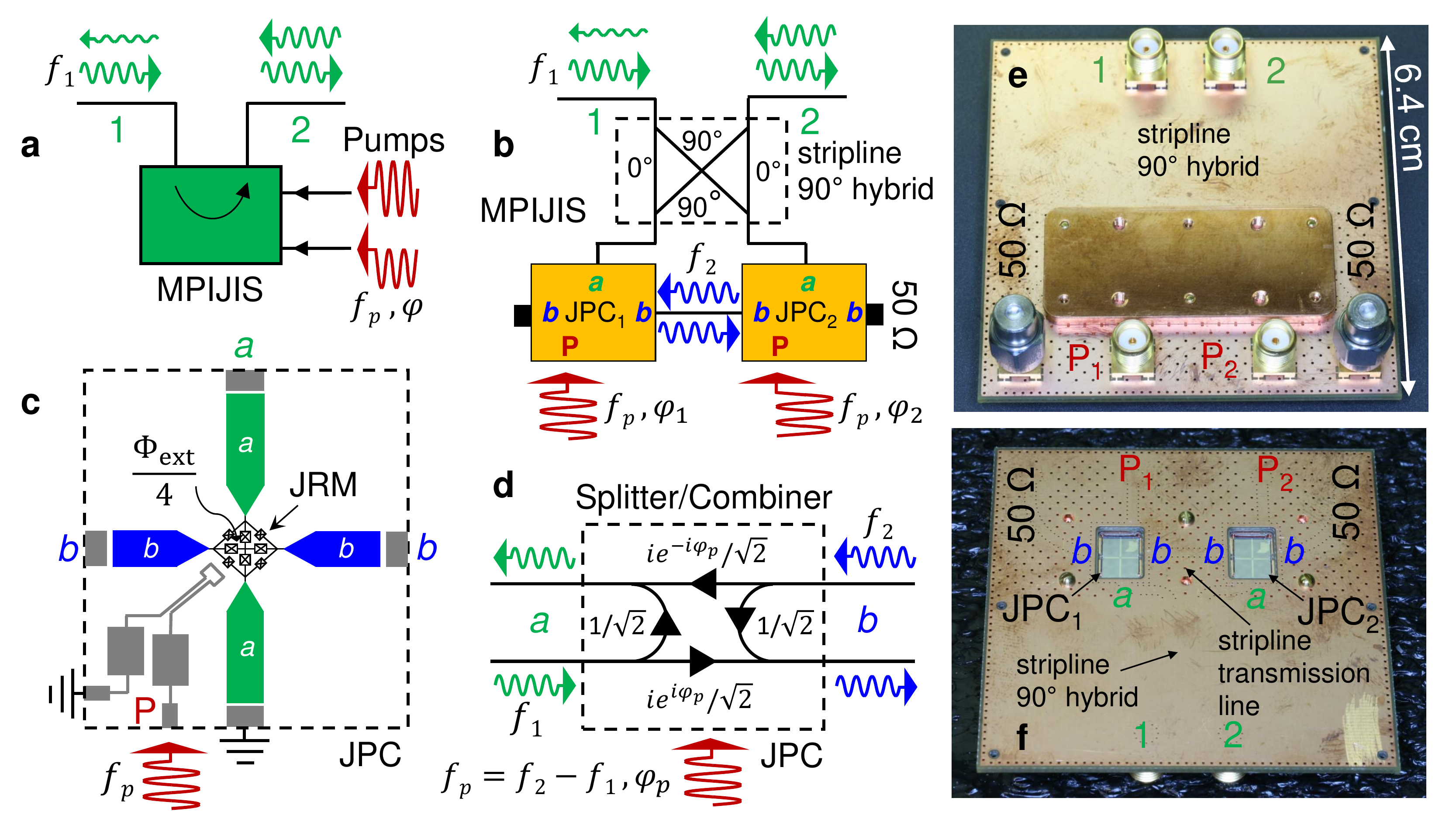}
		\caption{\textbf{a} Black-box illustration of the MPIJIS operation. The MPIJIS receives two microwave drives (pumps) at the same frequency $f_p$ having a certain balanced amplitude and  phase difference $\varphi=-\pi/2$. The MPIJIS transmits signals within the device bandwidth, e.g., $f_1$, from port 1 to 2 with almost unity transmission, while significantly attenuating signals propagating in the opposite direction. \textbf{b} Block circuit diagram of the MPIJIS measured in this work. The device consists of two Josephson parametric converters (JPCs), i.e., nondegenerate Josephson mixers, whose ports \textit{a} and \textit{b} are coupled by a $90^{\circ}$ hybrid and a short transmission line, respectively, which form an interferometric setup for propagating signals at frequencies $f_1$ and $f_2$ supported by the JPCs. One side of port \textit{b} of each JPC is terminated with a matched $50$ $\Omega$ cold load forming the internal ports of the device. The pump is fed to each JPC through a separate physical port. \textbf{c} Schematic layout of the JPC employed in the MPIJIS. The JPC consists of two half-wavelength microstrip resonators, denoted as \textit{a} and \textit{b}, strongly coupled to a Josephson ring modulator (JRM), which functions as a dispersive, three-wave-mixing medium. Each resonator is capacitively coupled to one external feedline on either side via gap capacitances. In this configuration, resonator \textit{a} is made single ended by shorting one of its feedlines to ground. The pump is injected to the JRM through an on-chip drive line. \textbf{d} Signal flow graph for the JPC operated at a $50:50$ beam splitter/combiner working point, in which half of the input signals are reflected off and the other half is transmitted to the other port with frequency conversion. The transmitted signals acquire a nonreciprocal phase shift based on the pump phase. \textbf{e} and \textbf{f} exhibit photos of the top and bottom view of the MPIJIS, respectively, realized by embedding two JPC chips into a multilayer printed circuit board, which incorporates a stripline $90^{\circ}$ hybrid and transmission lines (the center conductor of the stripline components is buried and therefore not visible). In \textbf{f}, the bottom cover that encloses the chips is removed to expose the JPCs. The device photos \textbf{e} and \textbf{f} are adapted from Ref. \cite{JDAQST}.      
		}
		\label{Device}
	\end{center}
\end{figure*}
\begin{figure*}
	[tb]
	\begin{center}
		\includegraphics[
		width=1.7\columnwidth 
		]%
		{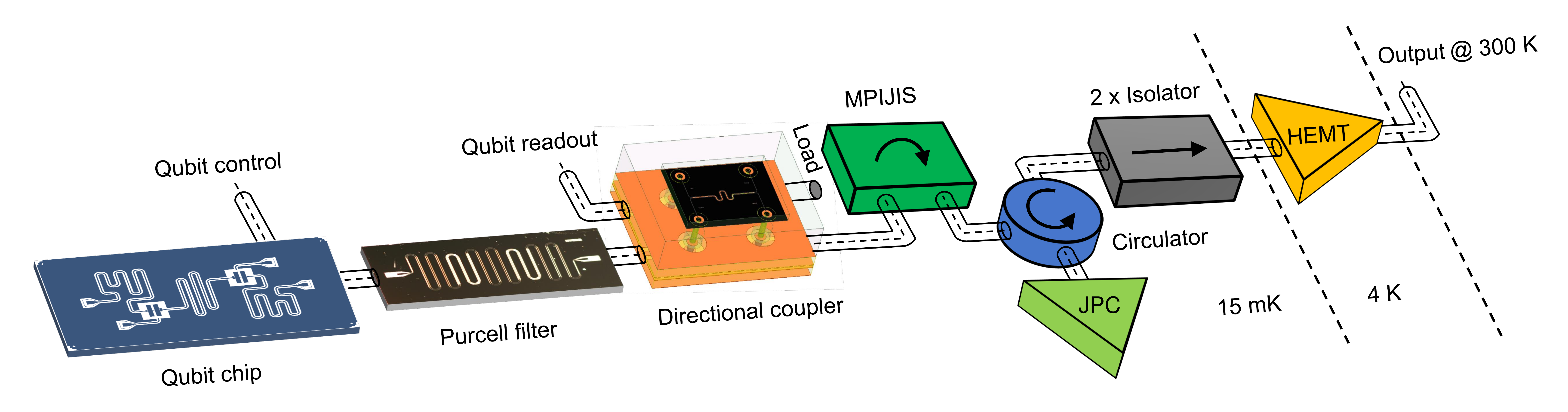}
		\caption{Rapid, high-fidelity, QND measurement setup of a superconducting qubit, which incorporates active isolation. The setup, which only exhibits the main components, includes a two-qubit chip in which only one qubit is measured. The measured qubit is read through a single-port, fast readout resonator and controlled via a separate line, which is capacitively coupled to the qubit. To prevent the qubit from relaxing through the readout resonator, a Purcell filter is added to the resonator port, which connects to the rest of the setup. To enable a reflective measurement of the qubit without utilizing a circulator, a broadband, on-chip directional coupler is incorporated between the Purcell filter and the readout input line. In this configuration, the readout signals input on the directional coupler couple, with about $20$ dB of attenuation, to the Purcell filter (and readout resonator), whereas the output readout signals reflecting off the readout resonator are transmitted, with a small insertion loss, to a MPIJIS incorporated into the output chain. Following the MPIJIS, a cryogenic circulator is used to route the readout signal to a quantum-limited Josephson amplifier working in reflection (i.e., JPC) and route the amplified reflected signal off the JPC towards the following amplification stage along the output chain, i.e., the HEMT, which is separated from the JPC circulator by two wideband cryogenic isolators, which further protect the qubit from noise coming down the output chain. 
		}
		\label{MeasSetup}
	\end{center}
\end{figure*}

\begin{figure*}
	[tb]
	\begin{center}
		\includegraphics[
		width=1.65\columnwidth 
		]%
		{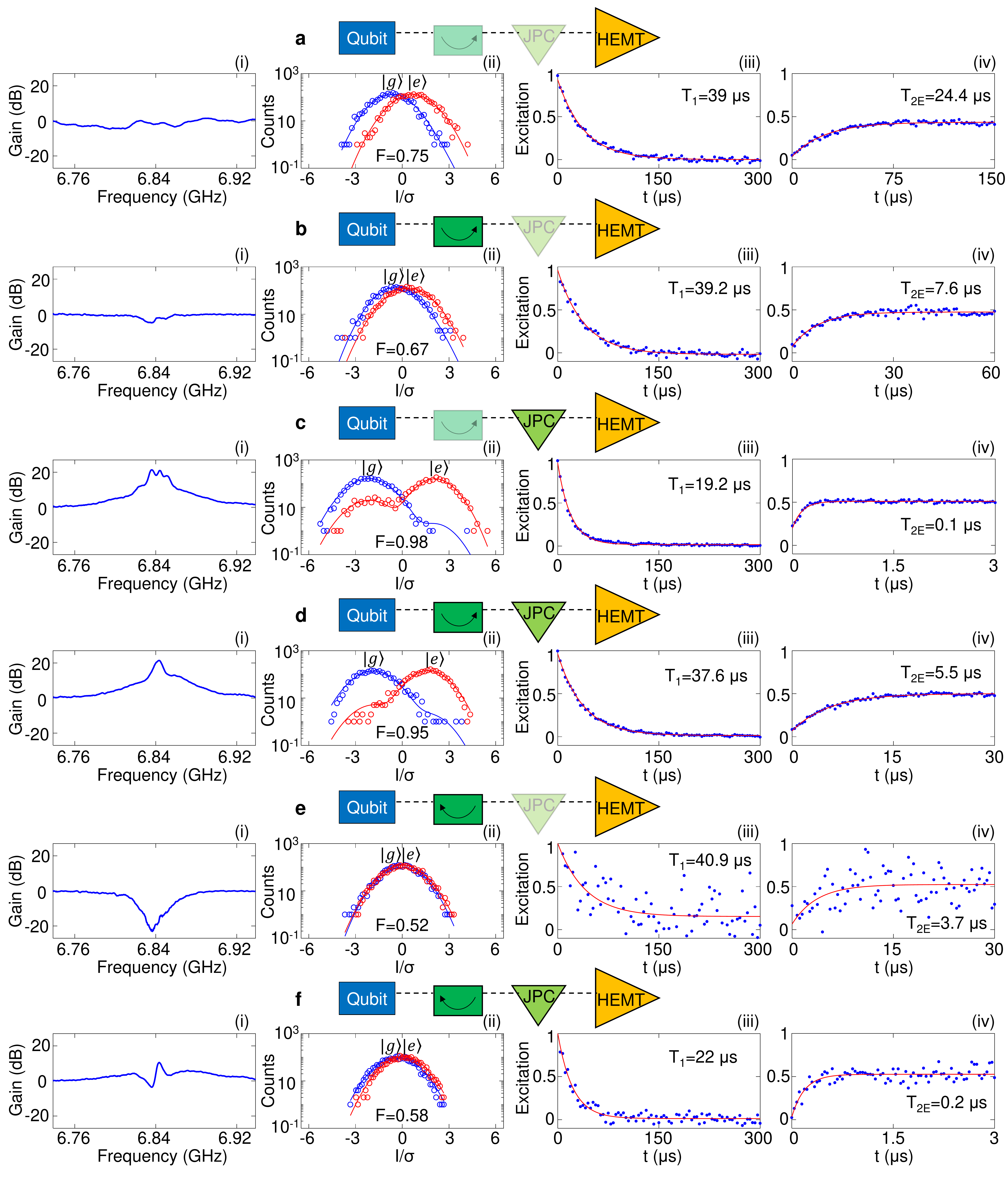}
		\caption{Panels \textbf{a}-\textbf{f} outline the six possible configurations of the high-fidelity, active-protection, qubit measurement setup shown in Fig.\,\ref{MeasSetup}. In \textbf{a}, the MPIJIS and JPC are \textit{off}. In \textbf{b}, the MPIJIS is \textit{on} and biased in the forward direction while the JPC is \textit{off}. In \textbf{c}, the MPIJIS is \textit{off} while the JPC is \textit{on}. In \textbf{d}, the MPIJIS and JPC are \textit{on} and the MPIJIS is biased in the forward direction. In \textbf{e}, the MPIJIS is \textit{on} and biased in the backward direction while the JPC is \textit{off}. In \textbf{f}, the MPIJIS and JPC are \textit{on} and the MPIJIS is biased in the backward direction. The measurement results exhibited in columns (i)-(iv) for each configuration correspond, respectively, to the net gain of the reflected readout signal (relative to the HEMT-only case), the qubit readout fidelity measurement, $T_1$ and $T_{\rm{2E}}$ of the qubit.                           
		}
		\label{CohFidSparam}
	\end{center}
\end{figure*}

\noindent \textbf{The device.} The MPIJIS realized in this work is shown in Fig.\,\ref{Device}. In Fig.\,\ref{Device}a, we show a black-box representation of the MPIJIS. It includes two external ports $1$ and $2$, which support propagating microwave signals at $f_1$, which fall within the device bandwidth. The MPIJIS receives, for its operation, two monochromatic microwave pumps at $f_p$ having a certain amplitude and phase difference $\varphi=\pm\pi/2$. The pumps are fed into the MPIJIS through auxiliary ports. In the example of Fig.\,\ref{Device}a, the MPIJIS is operated in the forward direction, i.e., propagating signals entering port $1$ (input) are transmitted in the direction of the arrow to port $2$ (output) with almost unity transmission, whereas propagating signals entering port $2$ are significantly attenuated upon exiting port $1$. Figure\,\ref{Device}b exhibits a block-circuit diagram of the MPIJIS. It consists of two nominally identical Josephson parametric converters (JPCs) \cite{JPCreview,JPCnature}, which serve as lossless, nondegenerate, three-wave mixing Josephson devices as outlined in Fig.\,\ref{IsolatorJISComp}d. The two JPCs are embedded into an interferometric setup, which couples modes \textit{a} and \textit{b} via a $90^{\circ}$ hybrid and a short transmission line, respectively. The pump drives are fed to the JPCs via separate physical ports (denoted as P) \cite{hybridLessJPC}. The circuit diagram also reveals two internal ports of the device coupled to ports \textit{b} and terminated by $50$ $\Omega$ loads. A schematic layout of the JPC is shown in Fig.\,\ref{Device}c. It consists of two half-wavelength, microstrip resonators \textit{a} and \textit{b}, which intersect in the center at a Josephson ring modulator (JRM) \cite{microstripJPC}. The JRM consists of four outer-loop Josephson junctions arranged in a Wheatstone bridge configuration. The four internal Josephson junctions inductively shunt the JRM and enable the resonance frequencies $\omega_{a,b}/2\pi$ of resonators \textit{a} and \textit{b} to be tuned as a function of the applied external magnetic flux $\Phi_{\rm{ext}}$ threading the JRM \cite{Roch}. When $0<|\Phi_{\rm{ext}}|<\Phi_0$, where $\Phi_0$ is the flux quantum, the JRM acts as a dispersive nonlinear medium mixing three orthogonal modes with a leading nonlinear term in the system Hamiltonian of the form $\mathcal{H}_{\rm{3wave}}={\hbar}g_3(\textbf{\textit{a}}+\textbf{\textit{a}}^{\dagger})(\textbf{\textit{b}}+\textbf{\textit{b}}^{\dagger})(\textbf{\textit{c}}+\textbf{\textit{c}}^{\dagger})$ \cite{JPCreview}. Here, $g_3$ is a flux-dependent coupling strength, \textbf{\textit{a}} and \textbf{\textit{b}} are the annihilation operators for the differential modes \textit{a} and \textit{b}, while \textbf{\textit{c}} is the annihilation operator for the mode \textit{c} common to both resonators. Each resonator is capacitively coupled via equal gap capacitors to two $50$ $\Omega$ feedlines, which carry incoming and outgoing signals. The photon decay rates $\kappa_{a,b}/2\pi\backsimeq40$ MHz of resonators \textit{a} and \textit{b} are primarily determined by the impedance mismatch created by the gap capacitors between the feedlines and resonators. In the MPIJIS configuration shown in Fig.\,\ref{Device}b, resonator \textit{a} is single ended (i.e., one feedline is shorted to ground) and the pump drive is directly injected to the JRM through a separate on-chip feedline as illustrated in Fig.\,\ref{Device}c \cite{hybridLessJPC}. When the JPC is operated as a frequency converter between modes \textit{a} and \textit{b}, a strong, coherent, off-resonant, common drive is applied at $\omega_c\equiv\omega_p=\omega_b-\omega_a$ \cite{Conv,QuantumNode}. With such a classical drive, we obtain in the rotating wave approximation $\mathcal{H}_{\rm{3wave}}={\hbar}|g_{ab}|(e^{i\varphi_p}\textbf{\textit{a}}\textbf{\textit{b}}^{\dagger}+e^{-i\varphi_p}\textbf{\textit{a}}^{\dagger}\textbf{\textit{b}})$, where $g_{ab}$ is a pump-amplitude-dependent coupling strength and $\varphi_p$ is the pump phase. On resonance, the transmission amplitude associated with this frequency conversion process is given by $t=2\rho/(1+\rho^2)$, where $\rho=|g_{ab}|/\sqrt{\kappa_a\kappa_b}$. The transmission amplitude varies between $0$ (no conversion) and $1$ (full conversion). Since this process is unitary, the reflection and transmission amplitudes satisfy the condition $|r|^2+|t|^2=1$. Another crucial property exhibited by $\mathcal{H}_{\rm{3wave}}$ is the pump-phase-dependent nonreciprocal phase shift imprinted on signals undergoing upconversion from mode \textit{a} to \textit{b} ($\varphi_p$) versus downconversion from \textit{b} to \textit{a} ($-\varphi_p$). This nonreciprocal phase shift is outlined in Fig.\,\ref{Device}d, which shows a signal flow graph for the JPC operated at a special frequency-conversion working point known as the $50:50$ beam splitter point, where $r=t=1/\sqrt{2}$.    

\begin{figure*}
	[tb]
	\begin{center}
		\includegraphics[
		width=1.6\columnwidth 
		]%
		{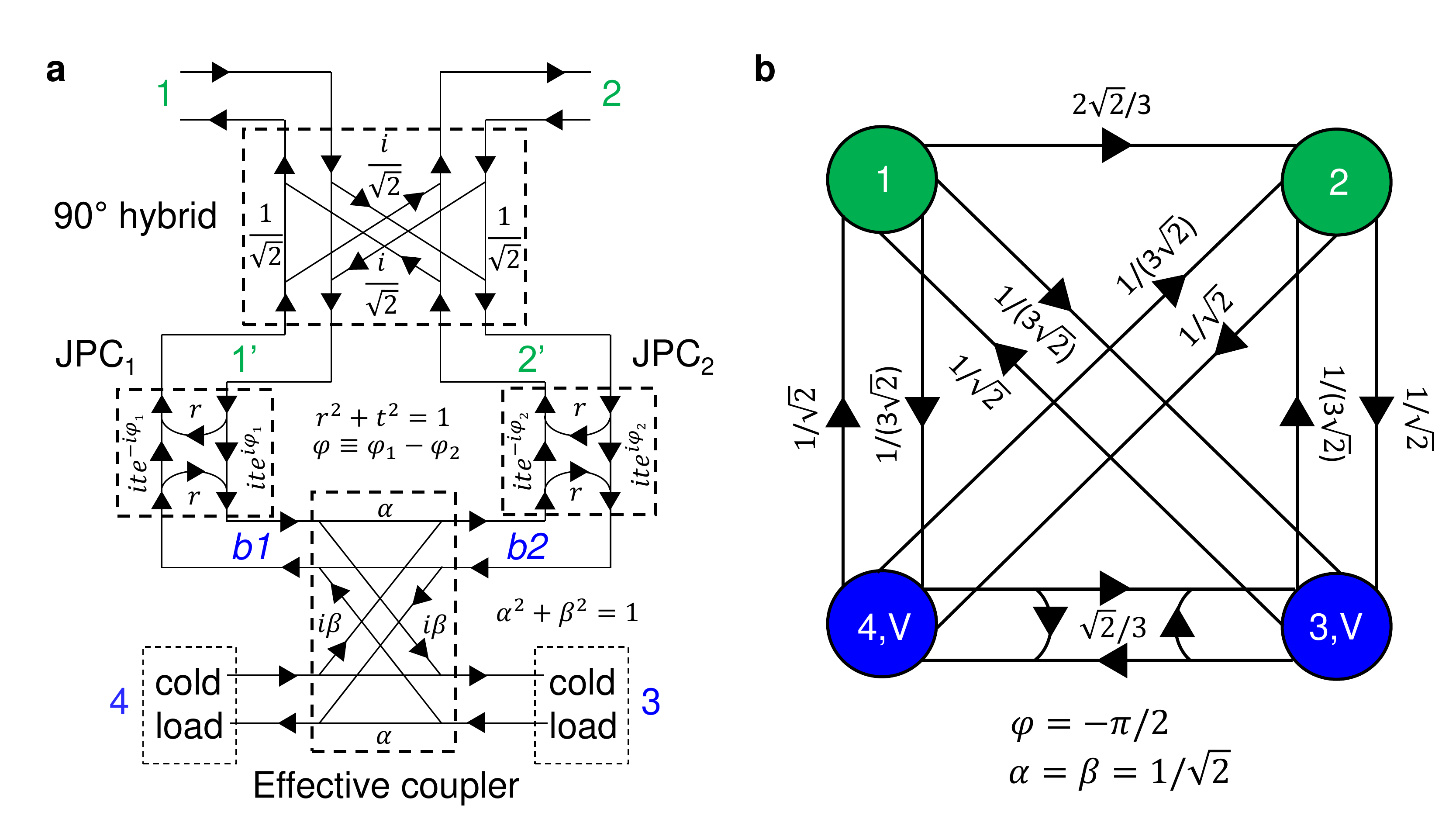}
		\caption{\textbf{a} Signal flow graph for the MPIJIS. It exhibits two JPCs operated in frequency conversion mode on resonance, i.e., $f_1=f_a$, $f_2=f_b$. Ports \textit{a} of the JPCs, denoted as $1^{\prime}$, $2^{\prime}$, are coupled via a $90^{\circ}$ hybrid while their internal \textit{b} ports, denoted as \textit{b1} and \textit{b2}, are coupled via a fictitious coupler, which models the attenuation present in the internal \textit{b} channel of the MPIJIS due to dissipation in the $50$ $\Omega$ cold terminations and the transmission line coupling the two JPCs. The coupler coefficients $\alpha$ and $\beta$ are taken to be real, satisfying the condition ${\alpha}^2+{\beta}^2=1$. The transmitted signals between ports \textit{a} and \textit{b} of the JPC undergo frequency conversion and acquire a nonreciprocal phase shift, which depends on the phase of the drive. \textbf{b} A graphical representation of the scattering parameters of the device for the case $r=t=1/\sqrt{2}$ (the JPCs are operated at the $50:50$ beam splitter working point), $\varphi=-\pi/2$, and the effective coupler is symmetrical and balanced, i.e., $\alpha=\beta=1/\sqrt{2}$.     
		}
		\label{SignalFlow}
	\end{center}
\end{figure*}

By coupling the two JPCs in an interferometric setup as shown in the block-circuit diagram depicted in Fig.\,\ref{Device}b (or device photos exhibited in Figs.\,\ref{Device}e, \ref{Device}f), we convert the nonreciprocal phase shift acquired by signals transversing the two JPCs, $\pm\varphi$, where $\varphi\equiv\varphi_1-\varphi_2$ is the phase gradient between the two pumps feeding the two mixers \cite{JPCgyrator}, into a nonreciprocal amplitude response via constructive or destructive wave-interference between multiple paths in the device. When solving the signal flow graph for the whole device (see Fig.\,\ref{SignalFlow} in the Methods section), we get on resonance the following transmission parameters for ports $1$ and $2$,    

\begin{equation} 
S_{2\leftrightarrows1}=i\frac{\sqrt{1-t^{2}%
	}\mp\sqrt{2}t^{2}\sin\varphi}{1+t^{2}},
\label{S12specialcase}
\end{equation}

\noindent and reflection parameters, 

\begin{equation} 
S_{11}=S_{22}=-i\frac{\sqrt{2}t^{2}\cos\varphi}{1+t^{2}}.
\label{S11specialcase}
\end{equation}

In the special case in which the JPCs are operated in the $50:50$ beam splitter point, i.e., $t=1/\sqrt{2}$, and the applied pump phase difference is $\varphi=-\pi/2$, we obtain, by substituting in Eqs.\,(\ref{S12specialcase}), (\ref{S11specialcase}), total isolation of signals transmitted from port $2$ to $1$, $S_{12}=0$, and vanishing reflections off ports $1$ and $2$, $S_{11}=S_{22}=0$. Whereas, for signals transmitted from port $1$ to $2$, we obtain almost unity transmission $|S_{21}|=2\sqrt{2}/3\backsimeq0.94$, which is equivalent to about $0.5$ dB loss in signal power. A detailed calculation of the scattering matrix of the MPIJIS as well as key measurement results are presented in the Methods section.    

It is worth noting that the same isolator circuit presented in Fig.\,\ref{Device}b can be used to realize a quantum-limited, phase-preserving Josephson directional amplifier, which has recently been demonstrated in Refs. \cite{JDA,JDAQST}. One main difference between the two interferometric nonreciprocal devices, relates to the mode of operation of the coupled JPCs. Here, the JPCs are operated in the frequency conversion mode with no photon gain \cite{Conv,JPCreview}, whereas in the directional amplifier application, they are operated in the nondegenerate amplification mode, where $\omega_p=\omega_a+\omega_b$ \cite{JPCreview,JPCnature,microstripJPC}. As a direct consequence of this difference, the two nonreciprocal devices differ in two important aspects, namely, added noise and stability. While the phase-preserving directional amplifier is required by quantum mechanics to add noise equivalent to a half input photon at the signal frequency $n_{\rm{add}}=1/2$, the added noise by the isolator is mainly set by its power attenuation in the forward direction $|S_{21}|^2$ \cite{Caves,QuantumNoiseIntro,Conv}. In the MPIJIS case, the added noise-equivalent-input-photons at the signal frequency is given by $n_{\rm{add}}=(1-|S_{21}|^2)/2|S_{21}|^2$ (see the Methods section). For an ideal MPIJIS, whose JPCs are joined by symmetric couplers and operated at the $50:50$ beam splitter point, we obtain $n_{\rm{add}}=1/16$ (corresponding to $|S_{21}|^2=8/9$ on resonance). Moreover, in the directional amplifier case, the amplitude-gain of each JPC is bounded by the amplitude-attenuation of the internal mode \textit{b} to ensure stability of the device in the presence of the feedback loop formed between the JPCs \cite{JDA,JDAQST}. Such a stability requirement is not applicable in the isolation case. 

It is also worth noting that the dual purpose of the present device, i.e., it can be operated as a directional amplifier or isolator by changing the pump frequency, amplitude, and phase is similar to other nonreciprocal Josephson devices reported recently \cite{ReconfJJCircAmpl,NRAumentado2}. 
	
\noindent \textbf{The quantum setup.} To demonstrate that the MPIJIS is suitable for quantum applications, we incorporate it into a high-fidelity qubit measurement setup as shown in Fig.\,\ref{MeasSetup}. The qubit chip used in the measurement consists of two coupled transmons, but only one transmon is measured. Each qubit has its own capacitively coupled control port, which is separate from the readout resonator port used for measurement. To enable fast readout, the qubit is dispersively coupled to a relatively large bandwidth readout resonator, in turn,  coupled to a superconducting Purcell filter integrated into the same PCB as the qubit chip (for further information see the Methods section). The Purcell filter is added to preserve the qubit lifetime by suppressing spontaneous emission of qubit excitations through the fast resonator. The effective readout bandwidth of the combined resonator-Purcell system is $\kappa/2\pi=7.99$ MHz. The qubit and readout frequencies are $\omega_q/2\pi=5.2696$ GHz and $\omega_r/2\pi=6.838$ GHz, respectively, while the qubit-state-dependent resonance frequency shift is $\chi/2\pi=3.4$ MHz. Another important novelty of this qubit setup is the integration of a custom-made, wideband, superconducting directional coupler to couple readout pulses into and out of the readout resonator. Compared to conventional qubit setups, which utilize a circulator to measure readout resonators in reflection, utilizing a directional coupler produces two key differences: (1) the input readout signal entering the directional coupler is attenuated by about $18$ dB, which can be effectively lumped into the total attenuation of the input readout line, (2) the directional coupler is reciprocal, therefore, does not protect the qubit against noise coming from the output chain. The main advantages of using a directional coupler over a magnetic circulator are compatibility with superconducting quantum circuits and integrability with other microwave components in the measurement scheme. Further information about the directional coupler performance, fabrication, and packaging can be found in the Methods section. Following the directional coupler, which transmits the output readout signal with minimal attenuation, we incorporate the MPIJIS and a quantum-limited Josephson amplifier (i.e., JPC), separated by a cryogenic circulator. The circulator is crucial, in this case, for three reasons: (1) it separates the weak readout signal input on the JPC from the amplified reflected output signal; (2) it preserves, to a large extent, the readout fidelity of the output signal due to its relatively low insertion loss ($<1$ dB); and (3) it partially protects the qubit (by about $15-18$ dB) against vacuum noise amplified in reflection by the JPC.  

\begin{figure*}
	[tb]
	\begin{center}
		\includegraphics[
		width=1.6\columnwidth 
		]%
		{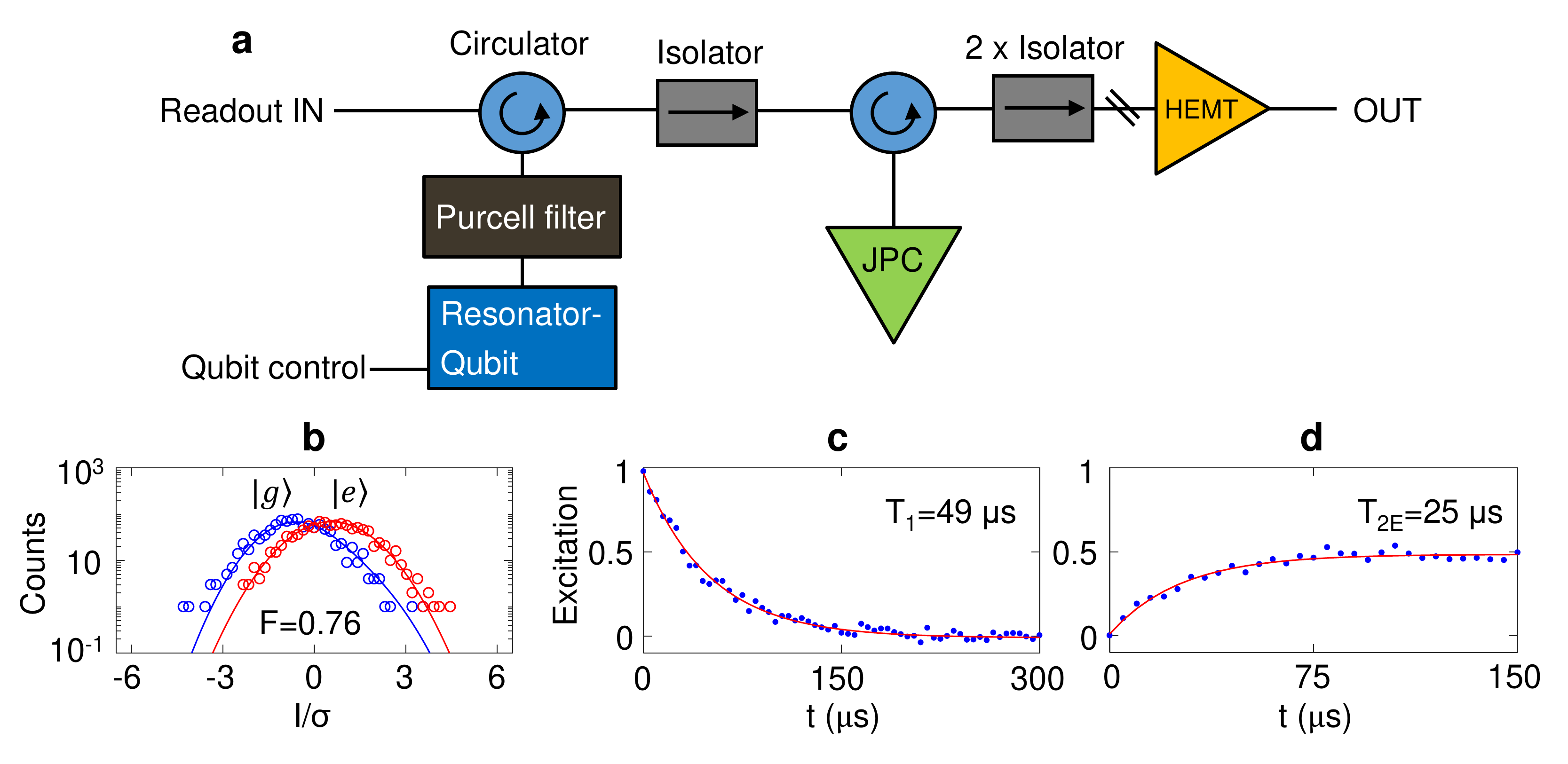}
		\caption{\textbf{a} Block-circuit diagram of a conventional rapid, high-fidelity qubit readout setup, which utilizes magnetic-based circulators and isolators. \textbf{b}, \textbf{c}, and \textbf{d} exhibit qubit measurements, i.e., qubit readout fidelity, $T_1$, and $T_{\rm{2E}}$, respectively, taken in a separate cooldown using the same qubit-resonator device and measurement parameters as Fig. \ref{CohFidSparam}. The JPC is turned off in these measurements.
		} 
		\label{TypHFmeas}
	\end{center}
\end{figure*}

\noindent \textbf{Qubit measurements.} Figure\,\ref{CohFidSparam} shows the main results of this work taken with the experimental setup of Fig.\,\ref{MeasSetup}. Figures\,\ref{CohFidSparam}a - \ref{CohFidSparam}f outline the six possible measurement configurations of the MPIJIS and the JPC. The graphs in columns (i)-(iv) exhibit measurement results taken for each configuration, namely, (i) normalized gain/attenuation of the MPIJIS and JPC versus frequency, (ii) readout fidelity histograms of the output chain extracted from the in-phase quadrature (I) of the output field corresponding to the qubit being initialized in the ground ($|g\rangle$) and excited ($|e\rangle$) states, (iii) relaxation time $T_1$, and (iv) decoherence time $T_2$ echo of the qubit. In configuration \textbf{a}, in which both the MPIJIS and JPC are \textit{off}, we obtain an almost flat transmission parameter because the MPIJIS is transparent for propagating signals and the JPC is totally reflective. Using this configuration, we obtain baseline values for the readout fidelity $0.75$, $T_1=39$ $\mu$s and $T_{\rm{2E}}=24.4$ $\mu$s. Very similar readout fidelities and coherence times are measured for this qubit when using the conventional high-fidelity setup instead, in which a cryogenic circulator and isolator replace the superconducting directional coupler and the MPIJIS, respectively (see Fig.\,\ref{TypHFmeas} in the Methods section). In configuration \textbf{b}, we operate the MPIJIS in the forward direction while keeping the JPC \textit{off}. In this case, we observe a small dip in the transmission near resonance of about $3-4$ dB, which can be attributed to a nonideal constructive interference taking place in the MPIJIS at this working point. The observed small dip also explains the slight decrease in the readout fidelity $0.67$ (ii). While we do not observe a decrease in the qubit lifetime due to the operation of the MPIJIS as seen in column (iii), we do observe a reduction, by about a factor of $3$, in $T_{\rm{2E}}=7.6$ $\mu$s, which we attribute to increased qubit dephasing as a result of pump-photons leaking into the readout resonator (see Fig.\,\ref{Spectrum} in the Methods section). In configuration \textbf{c}, we turn off the MPIJIS and turn on the JPC with a gain in excess of $20$ dB at the readout frequency and a dynamical bandwidth of about $20$ MHz as shown in column (i). As expected in this quantum-limited amplification scenario, the readout fidelity is significantly enhanced $F=0.98$. However, the enhancement in the fidelity is accompanied by a strong backaction on the qubit state, as seen in the drop of $T_1=19.2$ $\mu$s by about a factor of $2$ and the significant reduction of $T_{\rm{2E}}=0.1$ $\mu$s by about a factor of $200$, compared to configuration \textbf{a}. This backaction is mainly caused by certain amplified in-band quantum noise, which couple to the resonator-qubit system due to insufficient isolation (i.e., only one isolation stage is present between the qubit and JPC). Next, in configuration \textbf{d}, we keep the JPC \textit{on} and operate the MPIJIS in the forward direction. Remarkably, by turning on the active isolation, $T_1=37.6$ $\mu$s is almost restored to the baseline-value of configuration \textbf{a}, while $T_{\rm{2E}}=5.5$ $\mu$s is enhanced by a factor of $55$ compared to configuration \textbf{c} (i.e., without the active isolation). Moreover, the observed enhancements in the coherence times of the qubit compared to configuration \textbf{c} are achieved without a considerable impact on the total gain $20$ dB and readout fidelity $F=0.95$. In the last two remaining configurations \textbf{e} and \textbf{f}, the MPIJIS is operated in the backward direction, which is attained \textit{in situ} by shifting the pump phase difference of the two drives by $\pi$ compared to configurations \textbf{b} and \textbf{d}. In this mode of operation, the MPIJIS effectively mimics an attenuator of more than $20$ dB in the path of the readout signals as seen in column (i) of configuration \textbf{e} for which the JPC is \textit{off}. As expected in the case of heavy attenuation of configuration \textbf{e}, the readout fidelity is diminished $F=0.52$. Similar effect can also be seen in the poor signal to noise ratio of the coherence times measurements, which have a large scatter in the data points (taken with the same amount of averaging as the other cases). In general, the coherence times measured for the backward-operated isolator $T_1=40.9$ $\mu$s, $T_{\rm{2E}}=3.7$ $\mu$s are comparable to those of the forward-operated isolator (i.e., configuration \textbf{b}) within the measurement error margin. Finally, in configuration \textbf{f}, in which the JPC is \textit{on}, the amplification of the JPC almost cancels the attenuation of the MPIJIS as seen in (i), which leads to a slight enhancement of the readout fidelity $F=0.58$. The coherence times $T_1=22$ $\mu$s, $T_{\rm{2E}}=0.2$ $\mu$s measured for this case are quite similar to configuration \textbf{c}, because in both cases the isolator is not shielding the qubit against excess backaction of the preamplifier (JPC). 

To quantify the backaction effect on $T_{\rm{2E}}$ corresponding to the different configurations, we note that because $1/T_{\rm{2E}}=1/2T_1+1/T_{\phi}$ and $T_{\rm{2E}}<2T_1$, it follows that, in our case, the qubit coherence time is limited by dephasing, where $T_{\phi}$ is the dephasing time. One dominant dephasing mechanism in a qubit-resonator system operating in the dispersive coupling regime, such as ours, is photon shot noise in the resonator mode, which causes the qubit frequency to fluctuate due to the AC Stark effect \cite{AdamShotNoise,ChadShotNoise,CavityAtten,ClerkShotNoise}. In this case, the dephasing rate $\Gamma_{\phi}\equiv1/T_{\phi}$ is given by $\Gamma_{\phi}=\bar{n}\kappa\chi^2/(\kappa^2+\chi^2)$, where  $\bar{n}$ is the average photon number in the resonator, which can be written as $\bar{n}=\bar{n}_{\rm{th}}+\bar{n}_{\rm{ba}}$, where $\bar{n}_{\rm{th}}$ and $\bar{n}_{\rm{ba}}$ are the average thermal and nonthermal (backaction) photon numbers, respectively. Using these relations, we can evaluate $T_{\phi}$ and $\bar{n}_{\rm{ba}}$, summarized in Table \ref{table:1}, for the different configurations \textbf{a}-\textbf{f} exhibited in Fig.\,\ref{CohFidSparam}. We also extract  $\bar{n}_{\rm{th}}=0.004$ using configuration \textbf{a}, for which the MPIJIS and JPC are \textit{off} (no backaction is present). As expected, the largest values for $\bar{n}_{\rm{ba}}$ $1.29$ and $0.64$ are obtained for configurations \textbf{c} and \textbf{f} in which the JPC is \textit{on} while the MPIJIS is \textit{off} or operated in the backward direction (i.e., in a nonprotective mode). Conversely, a much smaller photon number by about two orders of magnitude $0.01-0.02$ is observed when the MPIJIS is operated in the protective forward direction.   

\begin{table}[h!]
	\centering
	\begin{tabular}{|c c c | c | c | c | c |} 
		\hline
		Config. & MPIJIS & JPC & $T_1$ ($\mu$s) & $T_{\rm{2E}}$ ($\mu$s) & $T_{\phi}$ ($\mu$s) & $\bar{n}_{\rm{ba}}$ \\ [0.5ex] 
		\hline\hline
		\textbf{a} & \textit{off} & \textit{off} & 39 & 24.2 & 35.1 & 0 \\ 
		\textbf{b} & \textit{F} & \textit{off} & 39.2 & 7.6 & 8.4 & 0.01 \\
		\textbf{c} & \textit{off} & \textit{on} & 19.2 & 0.1 & 0.1 & 1.29 \\
		\textbf{d} & \textit{F} & \textit{on} & 37.6 & 5.5 & 5.9 & 0.02\\
		\textbf{e} & \textit{B} & \textit{off} & 40.9 & 3.7 & 3.9 & 0.03 \\  
		\textbf{f} & \textit{B} & \textit{on} & 22 & 0.2 & 0.2 & 0.64 \\ 
		\hline
	\end{tabular}
	\caption{Evaluated $T_{\phi}$ and $\bar{n}_{\rm{ba}}$ for the different configurations. The symbols \textit{F} and \textit{B} correspond to the MPIJIS operated in the forward and backward direction, respectively.}
	\label{table:1}
\end{table}

\noindent \textbf{Discussion}

\begin{figure*}
	[tb]
	\begin{center}
		\includegraphics[
		width=1.6\columnwidth 
		]%
		{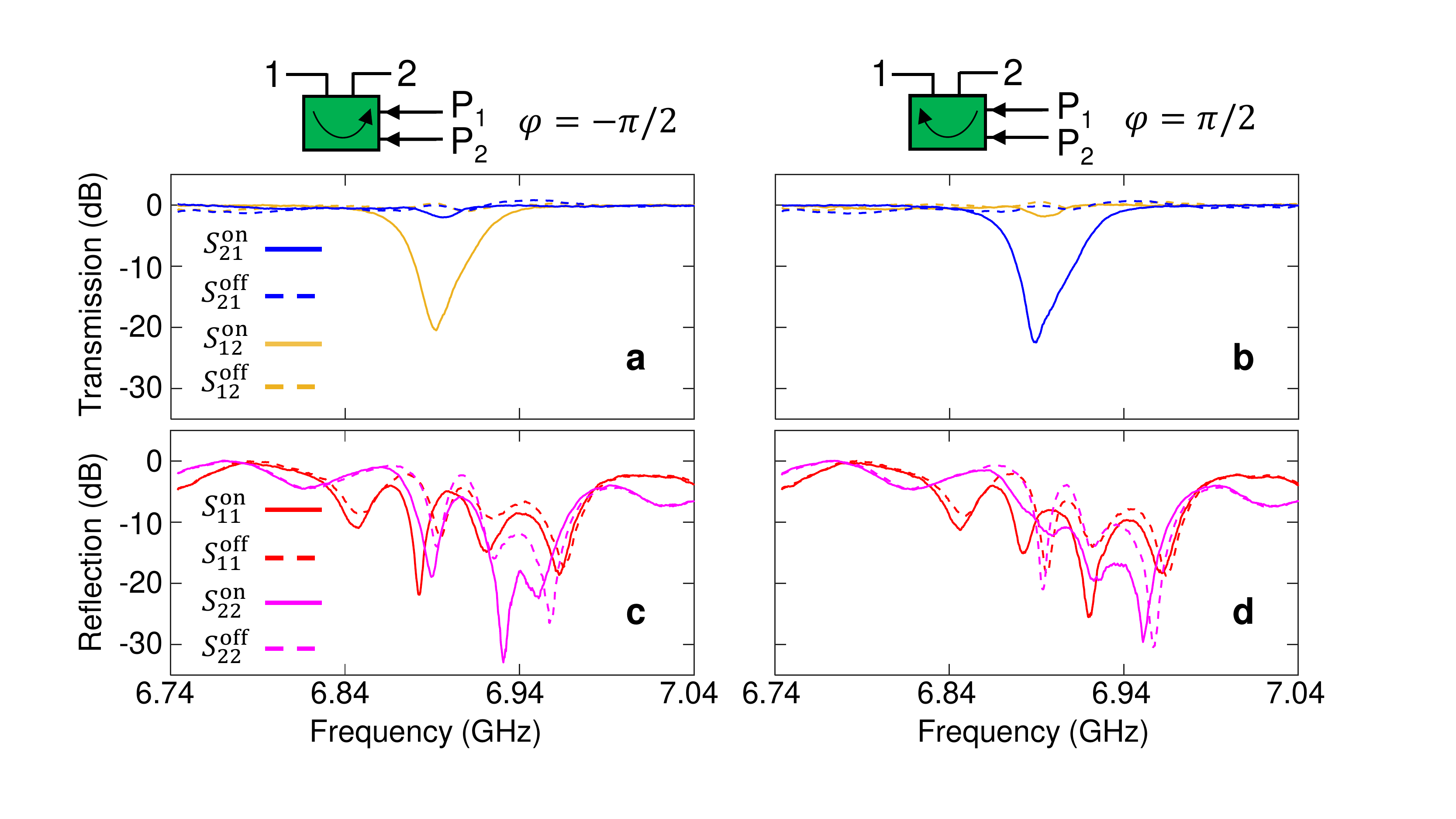}
		\caption{Measured scattering parameters of the MPIJIS versus frequency. \textbf{a} and \textbf{b} show the transmission parameters $|S_{21}|^2$ (solid blue) and $|S_{12}|^2$ (solid orange) measured for the MPIJIS biased in the forward ($\varphi=-\pi/2$) and backward ($\varphi=\pi/2$) direction, respectively. \textbf{c} and \textbf{d} show the corresponding reflection parameters $|S_{11}|^2$ (solid red) and $|S_{22}|^2$ (solid magenta). The dashed curves in \textbf{a}-\textbf{d} correspond to the MPIJIS being in the \textit{off} state. The applied pump frequency in this measurement is $2.8$ GHz.      
		}
		\label{SparamIsoMeas}
	\end{center}
\end{figure*}

\begin{figure}
	[tb]
	\begin{center}
		\includegraphics[
		width=\columnwidth 
		]%
		{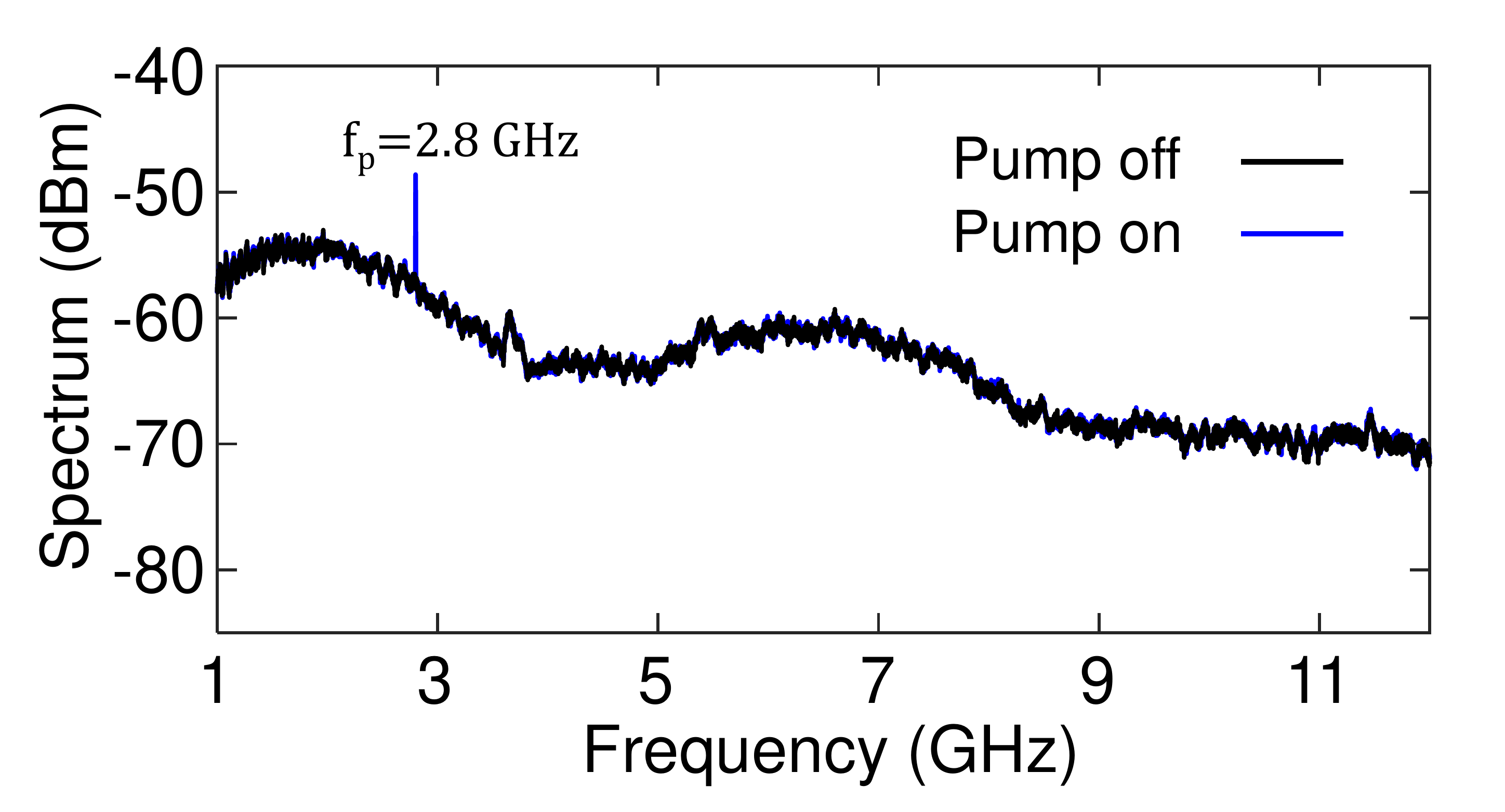}
		\caption{Output spectrum measurement of a MPIJIS port. The measurement is taken using a spectrum analyzer over a broad frequency range $1-11$ GHz. The black (blue) curve represents the spectrum measured at the output of the MPIJIS port, while the MPIJIS is off (on). The blue curve is measured while operating the MPIJIS at the same working point as in Fig.\,\ref{SparamIsoMeas}b, with the exception that in this case no input signal is applied to the device ports.               
		}
		\label{Spectrum}
	\end{center}
\end{figure}

\noindent Based on the results of Fig.\,\ref{CohFidSparam} and Fig.\,\ref{SparamIsoMeas} (in the Methods section), there are several areas in which the performance of this proof-of-principle MPIJIS can be further improved: (1) isolator-backaction, which is implied by the observed reduction of $T_{\rm{2E}}$ when the isolator is \textit{on} (Fig.\,\ref{CohFidSparam}b) versus \textit{off} (Fig.\,\ref{CohFidSparam}a). One likely explanation for this backaction is pump-photon-leakage from the on-chip pump lines into resonators \textit{a} of the JPCs as observed in the spectrum measurement of Fig.\,\ref{Spectrum} (see the Methods section). Such leakage can be eliminated by redesigning the basic lowpass-filtering element, which is part of the on-chip pump lines shown in Fig.\,\ref{Device}c, so that it blocks low-pump frequencies around $3$ GHz. The present filtering element, designed for a JPC-operation in the amplification mode \cite{hybridLessJPC}, has a cutoff frequency around $5$ GHz; (2) excess-insertion-loss for on-resonance transmitted signals in the range $1.8-4$ dB, as seen in Figs. \ref{CohFidSparam}b (i), \ref{SparamIsoMeas}a,b, which is higher than the $0.5$ dB  predicted for the ideal case (see device theory in the Methods section). This suggests that the constructive interference taking place in the forward direction is not optimal due to certain phase and amplitude imbalance of the PCB-hybrid \cite{JDAQST}. By substituting the present unoptimized PCB-hybrid with an optimized on-chip version, we believe this figure and the reflections off the device ports (see Figs. \ref{SparamIsoMeas}c,d) can be significantly reduced; and (3) narrow-bandwidth of the MPIJIS on the order of $11$ MHz, limited by the resonator bandwidths of the JPCs \cite{JPCreview,Conv}. One method which could be employed to enhance the device bandwidth is impedance-engineering of the JPC feedlines. Applying such a technique in the case of single-port Josephson parametric amplifiers has successfully yielded bandwidths in the range $0.6-0.7$ GHz \cite{JPAimpedanceEng,StrongEnvCoupling}, which correspond to more than $12$-fold enhancement compared to standard designs.   

Additional enhancements of the device include, (1) reducing its footprint by integrating all components on chip and using lumped-element realization of the JPCs \cite{LumpedJPC} and hybrids \cite{Lumpedhybrids}, (2) unifying the two external ports of the pumps, as shown in Figs. \ref{Device}e,f. This could be achieved by injecting a single-pump drive into the MPIJIS through an on-chip $90^{\circ}$ hybrid whose two output ports connect to the two-stage pump lines as proposed in Ref. \cite{JPCgyrator}.     

In conclusion, we have introduced and realized a new isolator device, which does not rely on magnetic materials or strong magnetic fields and is fully compatible with superconducting quantum circuits. The new isolator is comprised of two coupled, nondegenerate, three-wave Josephson mixers embedded in an interferometric scheme. The nonreciprocal response of the device is controlled by the phase gradient of the same-frequency microwave drives feeding the two Josephson mixers. Such a microwave-signal control could enable fast switching of the isolation direction on the fly with a time scale on the order of $15$~ns, which is mainly limited by the inverse dynamical bandwidth of the device. The realized isolator exhibits isolation in excess of $20$ dB, a dynamical bandwidth of $11$ MHz, insertion loss of about $-1.8$ dB in the forward direction, signal-power reflections off its input and output ports below $-10$ dB, and a maximum input power of $-108$ dBm. Furthermore, we have validated the applicability of this isolation scheme for quantum measurements by incorporating it into a superconducting qubit measurement setup, which includes a transmon coupled to a fast cavity, Purcell-filter, custom-made, broadband, superconducting directional coupler, and a quantum-limited, Josephson parametric amplifier. Using this novel setup, we have demonstrated fast, single-shot, high-fidelity, QND measurements of the quantum state while providing active protection of the qubit against amplified noise originated by the Josephson parametric amplifier. Owing to its numerous desired properties, an optimized version of this Josephson isolator may play a pivotal role in scalable quantum architectures.

\noindent\textbf{Methods}

\begin{figure*}
	[tb]
	\begin{center}
		\includegraphics[
		width=1.6\columnwidth 
		]%
		{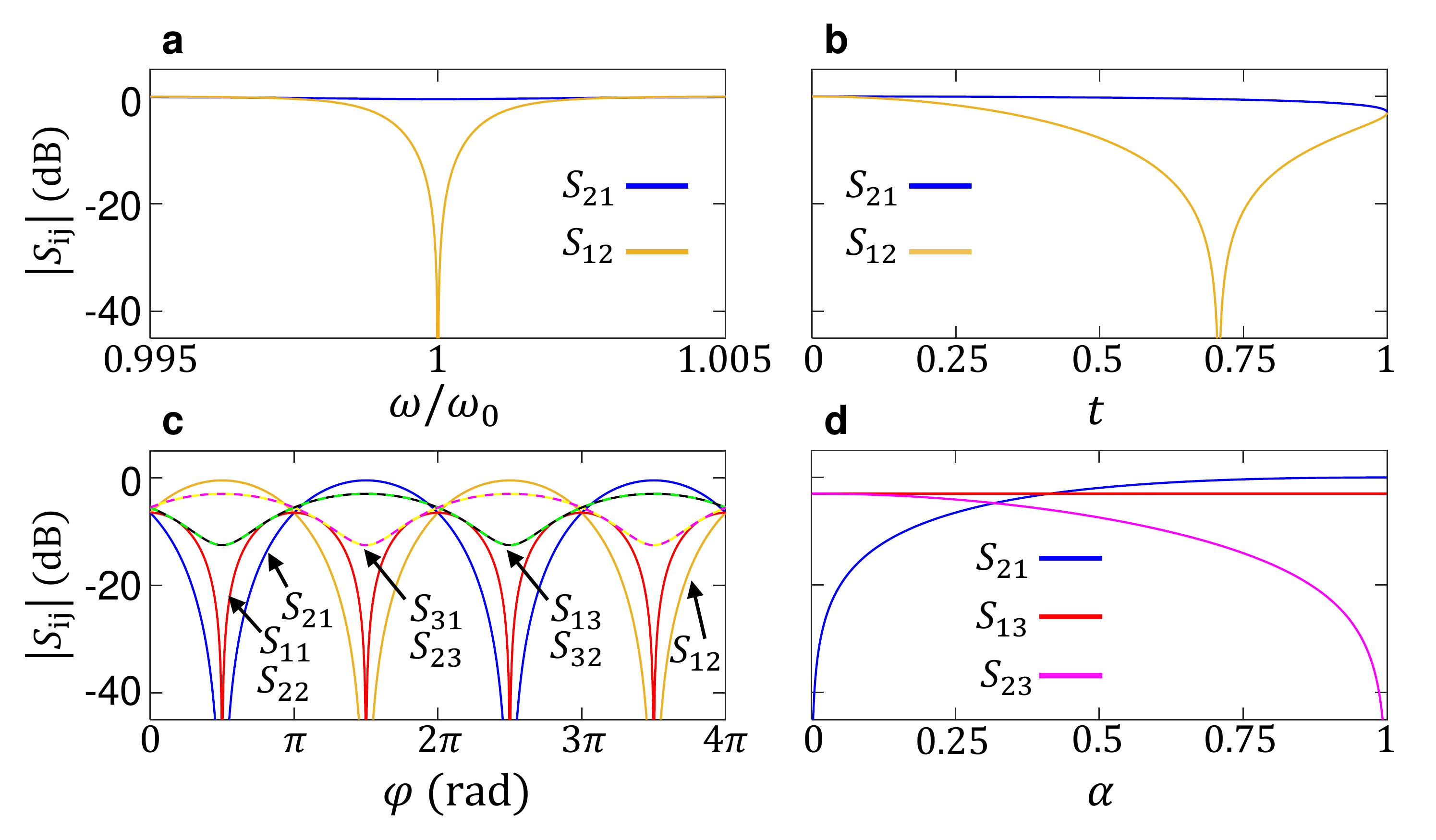}
		\caption{Theoretical results of the MPIJIS. \textbf{a} Magnitude of the transmission parameters $S_{21}$ (blue) $S_{12}$ (orange) versus normalized angular frequency $\omega/\omega_0$, where $\omega_0$ is the angular resonance frequency of resonator \textit{a}. \textbf{b} Magnitude of the transmission parameters $S_{21}$ (blue) $S_{12}$ (orange) on resonance versus the transmission parameter $t$ of the balanced JPCs. In both cases \textbf{a} and \textbf{b}, the MPIJIS is biased in the forward direction ($\varphi=-\pi/2$). \textbf{c} Periodic response of the various scattering parameters on resonance versus the pump phase difference $\varphi$. In the calculations \textbf{a} and \textbf{c}, the JPCs are operated at the $50:50$ beam splitter working point. In the calculations \textbf{a}, \textbf{b}, and \textbf{c}, the effective coupler is assumed to be a hybrid with equal real coefficients $\alpha=\beta=1/\sqrt{2}$. \textbf{d} $|S_{21}|^2$ (blue), $|S_{13}|^2$ (red), and $|S_{23}|^2$ (magenta) on resonance versus the effective coupler coefficient $\alpha$  which is varied between $0$ and $1$. In this calculation, $\beta=\sqrt{1-{\alpha}^2}$ and $t$ is calculated for each given $\alpha$ to yield a fixed isolation $|S_{12}|^2=0.01$ ($-20$ dB) on resonance.             
		}
		\label{ThyFig}
	\end{center}
\end{figure*}

\begin{figure}
	[tb]
	\begin{center}
		\includegraphics[
		width=\columnwidth 
		]%
		{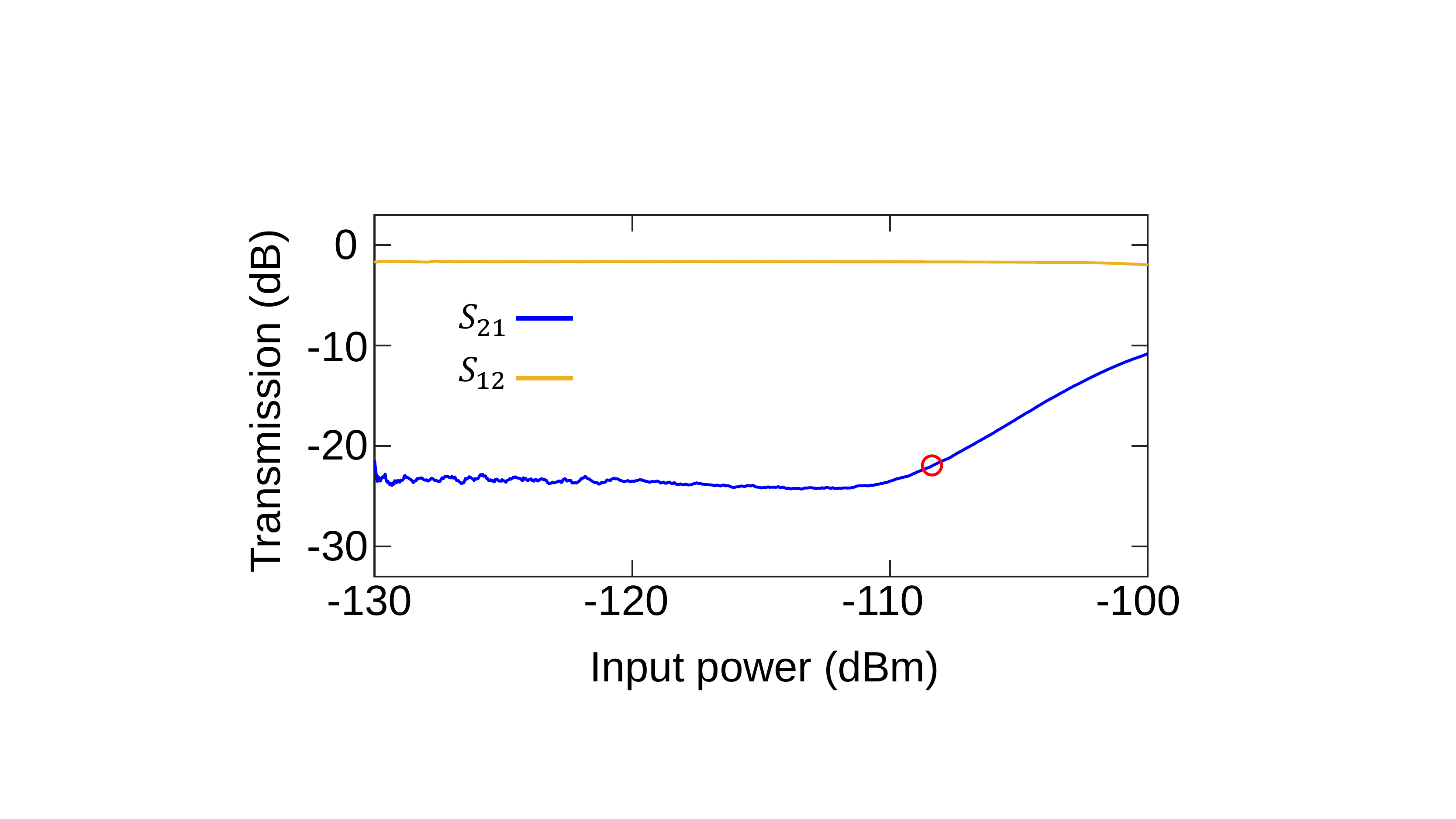}
		\caption{Maximum input power measurement of the MPIJIS taken on resonance. In this measurement, the MPIJIS is biased in the backward direction as shown in Fig.\,\ref{SparamIsoMeas}b. The red circle indicates the saturation power at which the isolation magnitude of the device represented by the blue line ($|S_{21}|^2$) decreases by $1$ dB compared to the low-power value. In comparison, $|S_{12}|^2$ represented by the orange line remains flat even beyond the saturation power.        
		}
		\label{DR}
	\end{center}
\end{figure}

\begin{figure*}
	[tb]
	\begin{center}
		\includegraphics[
		width=1.6\columnwidth 
		]%
		{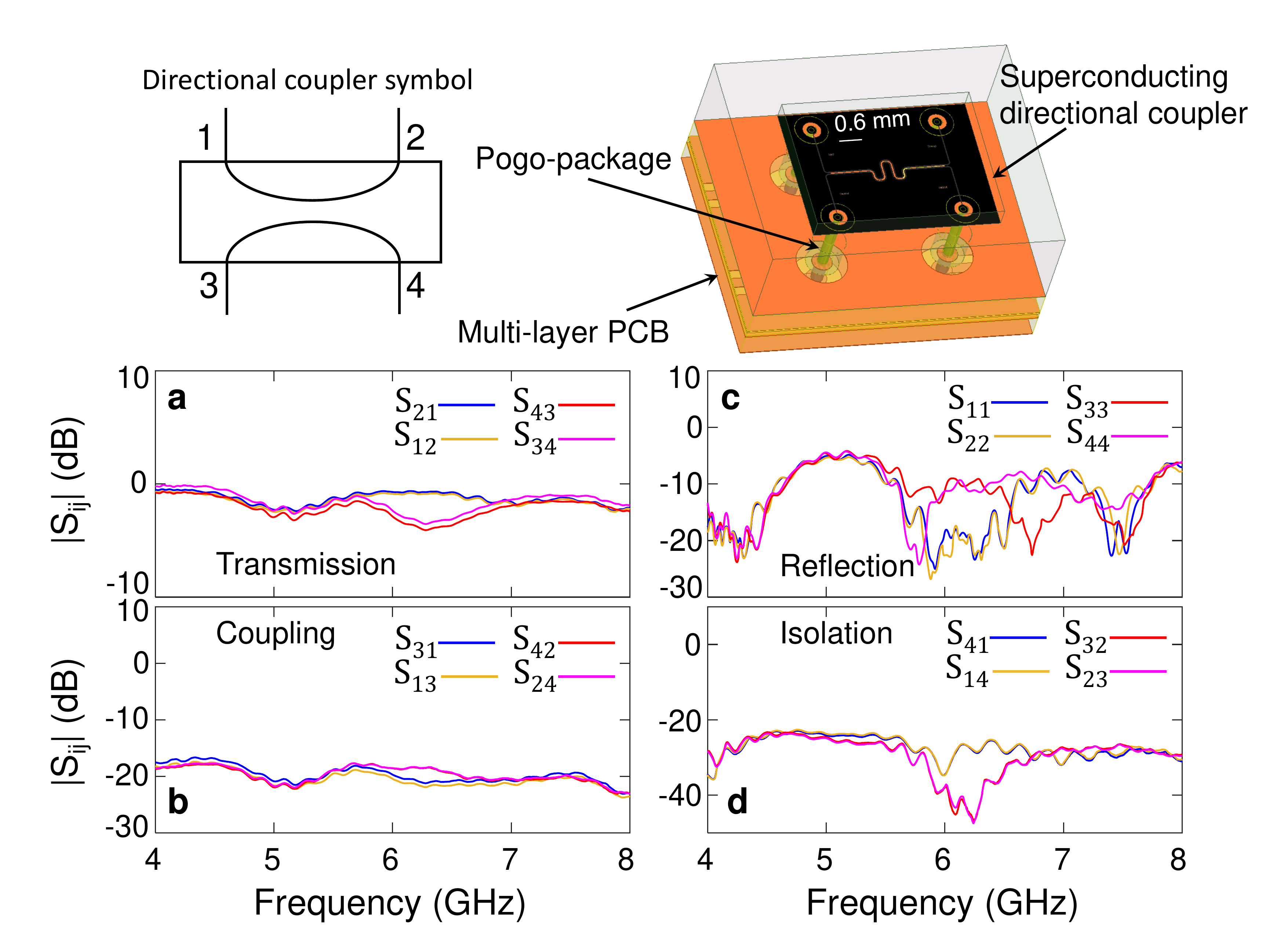}
		\caption{Measurement results for the scattering parameters of the on-chip, superconducting directional coupler used in the qubit measurement setup shown in Fig.\,\ref{MeasSetup}. The scattering parameters of the device are measured in a separate cooldown using four pairs of calibrated input and output lines, where each pair connects to one of the device ports shown in the top-left corner of the figure, through a three-port, commercial circulator (not drawn). In the top-right corner of the figure, we exhibit a schematic stack-up of the device, which consists of a superconducting directional coupler chip and multi-layer PCB which carries the incoming and outgoing signals via SMA connectors (not shown). The device also consists of a pogo-package which encloses the chip and connects between the directional coupler ports and designated metallic traces within the multi-layer PCB (see text and Fig.\,\ref{Pogo}). \textbf{a}-\textbf{d} show the magnitude of the transmission, coupling, reflection, and isolation parameters of the directional coupler, respectively, measured versus frequency between $4$ to $8$ GHz.              
		}
		\label{DircCoupRes}
	\end{center}
\end{figure*}

\begin{figure}
	[tb]
	\begin{center}
		\includegraphics[
		width=\columnwidth 
		]%
		{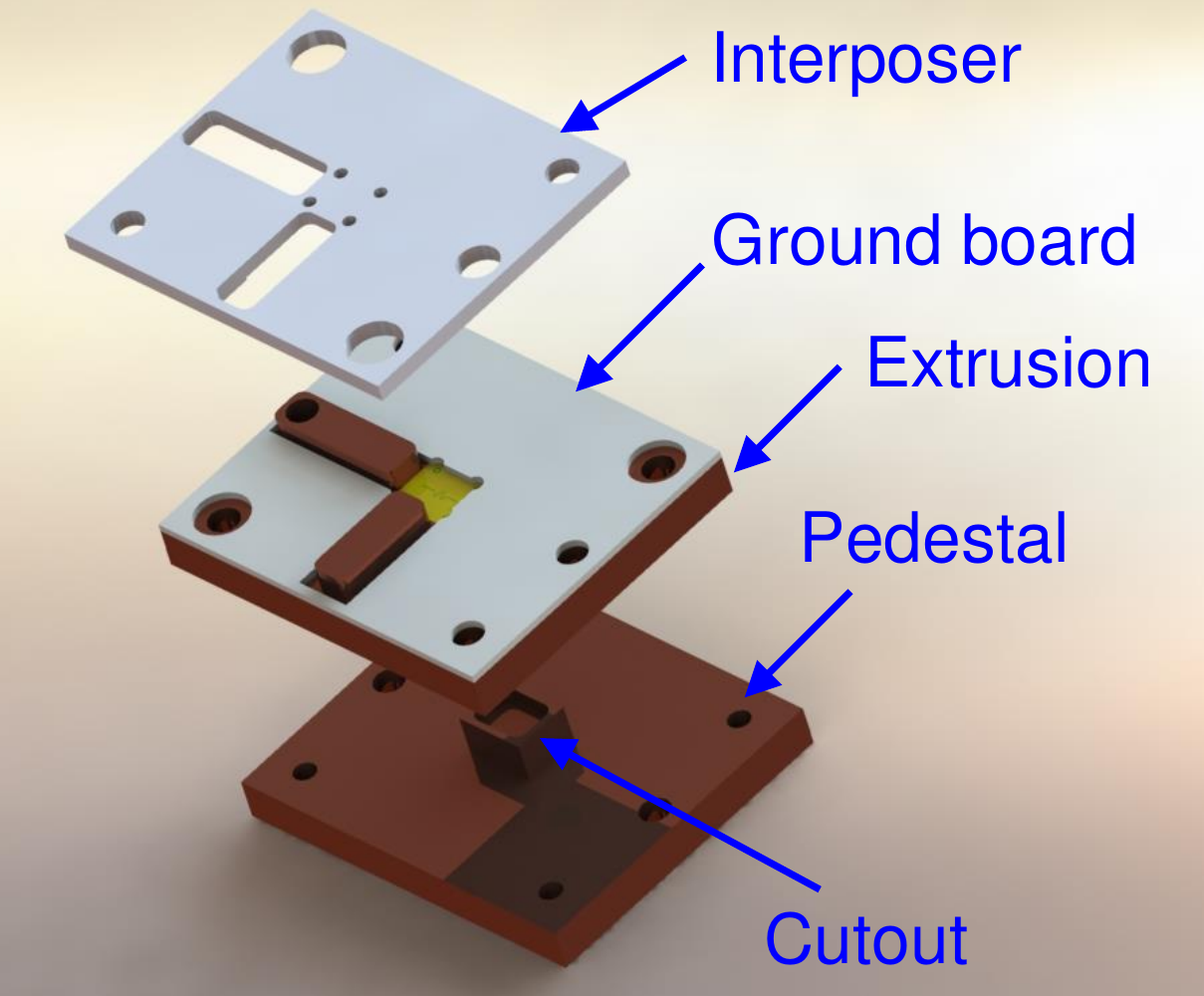}
		\caption{A schematic illustration of the pogo-package \cite{QST_Bronn2018}. It shows the different layers of the pogo-package used for housing the wideband, superconducting directional coupler. This package is used to provide $50$ $\Omega$ matched vertical transitions between the directional coupler chip and its PCB.                
		}
		\label{Pogo}
	\end{center}
\end{figure}

\noindent \textbf{MPIJIS theory.} To calculate the scattering parameters of the MPIJIS and demonstrate its isolation operation, we use the effective signal-flow graph exhibited in Fig.\,\ref{SignalFlow}a. The graph includes signal-flow graphs for the two coupled JPCs operated in frequency-conversion mode. On-resonance signals at $f_1=f_a$ or $f_2=f_b$ input on port \textit{a} (e.g., $1^{\prime}$ or $2^{\prime}$) or \textit{b} (e.g., \textit{b1} or \textit{b2}) are reflected off by a reflection-parameter $r$ and transmitted with frequency-conversion by a transmission-parameter $t$, where $r$ and $t$ are determined by the pump drive amplitude and satisfy the energy conservation condition $r^2+t^2=1$. In this calculation, we assume that the two JPCs are balanced, i.e., their reflection and transmission parameters are equal. It is worth noting that although we consider here the on-resonance case, it is straightforward to generalize the device response for signals that lie within the JPC dynamical bandwidth as we show below. The nonreciprocal phases $\varphi_1$ and $\varphi_2$ acquired by the frequency-converted transmitted signals between ports \textit{a} and \textit{b}, as indicated in the graph, correspond to the phases of the pump drives at frequency $f_p$ feeding $\rm{JPC_1}$ and $\rm{JPC_2}$, respectively. Figure\,\ref{SignalFlow}a also includes flow-graphs for two couplers coupling the \textit{a} and \textit{b} ports of the JPCs; one represents the $90^{\circ}$ hybrid, which couples between the \textit{a} ports of the JPCs, while the other is a fictitious one coup\-ling the \textit{b} ports. The main role of the latter coupler is to model the amplitude attenuation present on the \textit{b} port $\alpha$, due to signal absorption in the $50$ $\Omega$ cold loads and the insertion loss of the normal-metal transmission line coupling the two stages. Because of the structural symmetry of our device, we consider a symmetric coupler with real coefficients $\alpha$ and $\beta$, which satisfy the condition ${\alpha}^2+{\beta}^2=1$. For an ideal symmetric coupler (i.e., $90^{\circ}$ hybrid), $\alpha=\beta=1/\sqrt{2}$ \cite{Pozar}. 

In the stiff pump approximation, the JPC reflection and transmission parameter amplitudes can be written as \cite{JPCreview}

\begin{align}
\begin{array}
[c]{cc}%
r=\dfrac{1-\rho^2}{1+\rho^2}, \\
t=\dfrac{2\rho}{1+\rho^2},  
\end{array}
\label{r_s_res}%
\end{align}

\noindent where $0\leq\rho\leq1$ is a dimensionless pump amplitude. The lower bound $\rho=0$ corresponds to the case of no applied pump in which the JPC is \textit{off} and acts as a perfect mirror, whereas the upper bound $\rho=1$ corresponds to the case of full frequency conversion mode between ports \textit{a} and \textit{b}. By inspection \cite{Pozar}, the scattering matrix of the inner device defined by the ports $1^{\prime}, 2^{\prime}, 3, 4$, i.e., excluding the first $90^{\circ}$ hybrid, can be written in the form \cite{JDA}

\begin{align}
\left[  s\right]    & =\left(
\begin{array}
[c]{cccc}%
s_{1^{\prime}1^{\prime}} & s_{1^{\prime}2^{\prime}} & s_{1^{\prime}3} &
s_{1^{\prime}4}\\
s_{2^{\prime}1^{\prime}} & s_{2^{\prime}2^{\prime}} & s_{2^{\prime}3} &
s_{2^{\prime}4}\\
s_{31^{\prime}} & s_{32^{\prime}} & s_{33} & s_{34}\\
s_{41^{\prime}} & s_{42^{\prime}} & s_{43} & s_{44}%
\end{array}
\right)  \nonumber\\
& =\left(
\begin{array}
[c]{cccc}%
\frac{r\beta^{2}}{1-\alpha^{2}r^{2}} & -\frac{\alpha t^{2}e^{-i\varphi}%
}{1-\alpha^{2}r^{2}} & -\frac{\beta te^{-i\varphi_{1}}}{1-\alpha^{2}r^{2}} &
-\frac{\beta r\alpha te^{-i\varphi_{1}}}{1-\alpha^{2}r^{2}}\\
-\frac{\alpha t^{2}e^{i\varphi}}{1-\alpha^{2}r^{2}} & \frac{r\beta^{2}%
}{1-\alpha^{2}r^{2}} & -\frac{\beta r\alpha te^{-i\varphi_{2}}}{1-\alpha
	^{2}r^{2}} & -\frac{\beta te^{-i\varphi_{2}}}{1-\alpha^{2}r^{2}}\\
-\frac{\beta te^{i\varphi_{1}}}{1-\alpha^{2}r^{2}} & -\frac{\beta r\alpha
	te^{i\varphi_{2}}}{1-\alpha^{2}r^{2}} & -\frac{\beta^{2}r}{1-\alpha^{2}r^{2}}
& \frac{\alpha t^{2}}{1-\alpha^{2}r^{2}}\\
-\frac{\beta r\alpha te^{i\varphi_{1}}}{1-\alpha^{2}r^{2}} & -\frac{\beta
	te^{i\varphi_{2}}}{1-\alpha^{2}r^{2}} & \frac{\alpha t^{2}}{1-\alpha^{2}r^{2}}
& -\frac{\beta^{2}r}{1-\alpha^{2}r^{2}}%
\end{array}
\right),\label{s_inner_mat}%
\end{align}

\noindent where $\varphi\equiv\varphi_{1}-\varphi_{2}$. As we show below, it is this phase difference between the modulation phases of the two pumps feeding the two parametric active devices (i.e., the JPCs), which induces the nonreciprocal response of the MPIJIS \cite{NoiselessCirc,AhranovBohmPhotonic,AhranovBohmMixers,JPCgyrator}. The common coefficient $1/(1-\alpha^{2}r^{2})$ that appears in the scattering parameters of Eq.\,(\ref{s_inner_mat}) represents the sum over all possible reflections that the internal signals can experience in the self-loop formed between the two \textit{b} ports of the device. Unlike the directional amplification case \cite{JDA,JDAQST}, where the reflection-gain amplitude needs to be bounded to ensure stability, in the case of frequency conversion, with no photon gain, the scattering parameters of Eq.\,(\ref{s_inner_mat}) are stable for all values of $0\leq r\leq1$. In this simplified model, we assume that the phase acquired by signals at frequency $f_2$, propagating along the short transmission line between the two JPCs, is $2\pi k$ in each direction, where $k$ is an integer. In our device, the electrical length of the short transmission line is designed to give a phase of about $2\pi$ at $f_2$.  

It is straightforward to verify that the scattering matrix of Eq.\,(\ref{s_inner_mat}) is unitary (energy preserving). For example, it satisfies the condition

\begin{equation}
\left\vert s_{1^{\prime}1^{\prime}}\right\vert ^{2}+\left\vert s_{1^{\prime
	}2^{\prime}}\right\vert ^{2}+\left\vert s_{1^{\prime}3}\right\vert
^{2}+\left\vert s_{1^{\prime}4}\right\vert ^{2}=1.
\end{equation} 

Next, we derive the scattering matrix for the whole device, defined by ports $1$, $2$, $3$, $4$, which take into account the signal flow through the $90^{\circ}$ hybrid, 

\begin{align}
\left[  S\right]  & =
\left(
\begin{array}
[c]{cccc}%
S_{11} & S_{12} & S_{13} & S_{14}\\
S_{21} & S_{22} & S_{23} & S_{24}\\
S_{31} & S_{32} & S_{33} & S_{34}\\
S_{41} & S_{42} & S_{43} & S_{44}%
\end{array}
\right),
\label{S_mat_MPIJIS}
\end{align}

\noindent whose matrix elements are given by,

\begin{equation} 
S_{11}=\frac{1}{2}\left(  s_{1^{\prime}1^{\prime}}-s_{2^{\prime}2^{\prime}%
}+is_{2^{\prime}1^{\prime}}+is_{1^{\prime}2^{\prime}}\right), 
\label{params11}
\end{equation}
\begin{equation} 
 S_{12}=\frac{1}{2}\left(  is_{1^{\prime}1^{\prime}}+is_{2^{\prime}2^{\prime}}+s_{1^{\prime}2^{\prime}}-s_{2^{\prime}1^{\prime}}\right),
\label{params12}
\end{equation}
\begin{equation} 
S_{21}=\frac{1}{2}\left(  is_{1^{\prime}1^{\prime}}+is_{2^{\prime}2^{\prime}}+s_{2^{\prime}1^{\prime}}-s_{1^{\prime}2^{\prime}}\right), 
\label{params21}
\end{equation}
\begin{equation} 
S_{22}=\frac{1}%
{2}\left(  s_{2^{\prime}2^{\prime}}-s_{1^{\prime}1^{\prime}}+is_{2^{\prime}1^{\prime}}+is_{1^{\prime}2^{\prime}}\right),
\label{params22}
\end{equation}

\begin{align}
\begin{array}
[c]{cc}%
S_{13}=\frac{1}{\sqrt{2}}\left(
is_{2^{\prime}3}+s_{1^{\prime}3}\right), &
S_{14}=\frac{1}{\sqrt{2}}\left(
is_{2^{\prime}4}+s_{1^{\prime}4}\right),\\
S_{23}=\frac{1}{\sqrt{2}}\left(
s_{2^{\prime}3}+is_{1^{\prime}3}\right),&
S_{24}=\frac{1}{\sqrt{2}}\left(
s_{2^{\prime}4}+is_{1^{\prime}4}\right),
\end{array}
\label{params2}%
\end{align}

\begin{align}
\begin{array}
[c]{cc}%
S_{31}=\frac{1}{\sqrt{2}}\left(is_{32^{\prime}}+s_{31^{\prime}}\right), &
S_{32}=\frac{1}{\sqrt{2}}\left(s_{32^{\prime}}+is_{31^{\prime}}\right),\\
S_{41}=\frac{1}{\sqrt{2}}\left(is_{42^{\prime}}+s_{41^{\prime}}\right),&
S_{42}=\frac{1}{\sqrt{2}}\left(s_{42^{\prime}}+is_{41^{\prime}}\right),
\end{array}
\label{params3}%
\end{align}
 
\begin{align}
\begin{array}
[c]{cc}%
S_{33}=s_{33}, &
S_{34}=s_{34},\\
S_{43}=s_{43},&
S_{44}=s_{44}.
\end{array}
\label{params4}%
 \end{align}
 
Note that $[S]$ is unitary because $[s]$ is unitary and the $90^{\circ}$ hybrid is a unitary device. One prominent property seen from Eqs.\,(\ref{params11})-(\ref{params4}), is the interferometric nature of the device, manifested in its scattering parameters, which represent the sum over all possible paths that the waves can propagate in it.   

By substituting the scattering parameters of the inner device listed in Eq.\,(\ref{s_inner_mat}) into Eqs.\,(\ref{params11})-(\ref{params4}), and by writing the resulting expressions in terms of the parameter $t$, we obtain the scattering parameters of the MPIJIS in an explicit form 

\begin{equation}
S_{21}=\frac{i}{1+\frac{\alpha^{2}}{\beta^{2}}t^{2}}\left[  \sqrt{1-t^{2}%
}-\frac{\alpha}{\beta^{2}}t^{2}\sin\varphi\right]  ,\label{S21}\\
\end{equation}

\begin{equation}
S_{12}=\frac{i}{1+\frac{\alpha^{2}}{\beta^{2}}t^{2}}\left[  \sqrt{1-t^{2}%
}+\frac{\alpha}{\beta^{2}}t^{2}\sin\varphi\right]  ,\label{S12}\\
\end{equation}

\begin{equation}
S_{11}=S_{22}=-\frac{i\alpha}{\beta^{2}}\frac{t^{2}}{1+\frac{\alpha^{2}%
	}{\beta^{2}}t^{2}}\cos\varphi,\label{S11}\\
\end{equation}

\begin{equation}
S_{33}=S_{44}=-\frac{\sqrt{1-t^{2}}}{1+\frac{\alpha^{2}}{\beta^{2}}t^{2}%
},\label{S33}\\
\end{equation}

\begin{equation}
S_{34}=S_{43}=\frac{\alpha}{\beta^{2}}\frac{t^{2}}{1+\frac{\alpha^{2}%
	}{\beta^{2}}t^{2}},\label{S34}\\
\end{equation}

\begin{equation}
S_{13}=-\frac{te^{-i\varphi_{s}/2+i\pi/4}%
}{\sqrt{2}\beta\left(1+\frac{\alpha^{2}}{\beta^{2}}t^{2}\right)  }\left[
\sqrt{1-t^{2}}\alpha e^{i\frac{\varphi}{2}+i\frac{\pi}{4}}+e^{-i\frac
	{\varphi}{2}-i\frac{\pi}{4}}\right]  ,\label{S13}\\
\end{equation}

\begin{equation}
S_{14}=-\frac{te^{-i\varphi_{s}/2+i\pi/4}%
}{\sqrt{2}\beta\left(1+\frac{\alpha^{2}}{\beta^{2}}t^{2}\right)  }\left[e^{i\frac{\varphi}{2}+i\frac{\pi}{4}}+\sqrt{1-t^{2}}\alpha e^{-i\frac
	{\varphi}{2}-i\frac{\pi}{4}}\right]  ,\label{S14}\\
\end{equation}

\begin{equation}
S_{23}=-\frac{te^{-i\varphi_{s}/2+i\pi/4}%
}{\sqrt{2}\beta\left(1+\frac{\alpha^{2}}{\beta^{2}}t^{2}\right)  }\left[
\sqrt{1-t^{2}}\alpha e^{i\frac{\varphi}{2}-i\frac{\pi}{4}}+e^{-i\frac
	{\varphi}{2}+i\frac{\pi}{4}}\right]  ,\label{S23}\\
\end{equation}

\begin{equation}
S_{24}=-\frac{te^{-i\varphi_{s}/2+i\pi/4}%
}{\sqrt{2}\beta\left(1+\frac{\alpha^{2}}{\beta^{2}}t^{2}\right)  }\left[
e^{i\frac{\varphi}{2}-i\frac{\pi}{4}}+\sqrt{1-t^{2}}\alpha e^{-i\frac
	{\varphi}{2}+i\frac{\pi}{4}}\right]  ,\label{S24}\\
\end{equation}

\begin{equation}
S_{31}=-\frac{te^{i\varphi_{s}/2+i\pi/4}%
}{\sqrt{2}\beta\left(1+\frac{\alpha^{2}}{\beta^{2}}t^{2}\right)  }\left[
\sqrt{1-t^{2}}\alpha e^{-i\frac{\varphi}{2}+i\frac{\pi}{4}}%
+e^{i\frac{\varphi}{2}-i\frac{\pi}{4}}\right]  ,\label{S31}\\
\end{equation}

\begin{equation}
S_{32}=-\frac{te^{i\varphi_{s}/2+i\pi/4}%
}{\sqrt{2}\beta\left(1+\frac{\alpha^{2}}{\beta^{2}}t^{2}\right)  }\left[
\sqrt{1-t^{2}}\alpha e^{-i\frac{\varphi}{2}-i\frac{\pi}{4}}%
+e^{i\frac{\varphi}{2}+i\frac{\pi}{4}}\right]  ,\label{S32}\\
\end{equation}

\begin{equation}
S_{41}=-\frac{te^{i\varphi_{s}/2+i\pi/4}%
}{\sqrt{2}\beta\left(1+\frac{\alpha^{2}}{\beta^{2}}t^{2}\right)  }\left[
e^{-i\frac{\varphi}{2}+i\frac{\pi}{4}}+\sqrt{1-t^{2}}\alpha
e^{i\frac{\varphi}{2}-i\frac{\pi}{4}}\right]  ,\label{S41}\\
\end{equation}

\begin{equation}
S_{42}=-\frac{te^{i\varphi_{s}/2+i\pi/4}%
}{\sqrt{2}\beta\left(1+\frac{\alpha^{2}}{\beta^{2}}t^{2}\right)  }\left[
e^{-i\frac{\varphi}{2}-i\frac{\pi}{4}}+\sqrt{1-t^{2}}\alpha
e^{i\frac{\varphi}{2}+i\frac{\pi}{4}}\right]  ,\label{S42}
\end{equation}

\noindent where $\varphi_{s}\equiv\varphi_{1}+\varphi_{2}$. In what follows, we examine a few special cases of interest and outline a few important properties of the device. 

Without applied pump, i.e., $t=0$, the MPIJIS scattering matrix reduces into

\begin{equation}
\left[  S\right]  =\left(
\begin{array}
[c]{cccc}%
0 & i & 0 & 0\\
i & 0 & 0 & 0\\
0 & 0 & -1 & 0\\
0 & 0 & 0 & -1
\end{array}
\right) \cdot\label{S_mat_t_zero}
\end{equation}

This result shows that when the device is \textit{off}, the MPIJIS is transparent for propagating signals and effectively behaves as a lossless transmission line with an added reciprocal phase shift of $\pi/2$ for transmitted signals within the bandwidth of the $90^{\circ}$ hybrid.
In the special case, where the MPIJIS is \textit{on} and the phase difference between the pumps is $\varphi=-\pi/2$, the scattering matrix can be written in the form

\begin{equation}
\left[  S\right]  =\left(
\begin{array}
[c]{cccc}%
0 & i\Delta   & -\sqrt{\frac{1-\Delta^{2}}{2}} &
-\sqrt{\frac{1-\Delta^{2}}{2}}\\
i\Sigma   & 0 & -i\sqrt{\frac{1-\Sigma^{2}}{2}} &
i\sqrt{\frac{1-\Sigma^{2}}{2}}\\
-\sqrt{\frac{1-\Sigma^{2}}{2}} & -i\sqrt{\frac{1-\Delta^{2}}{2}} & -\frac{\Sigma+\Delta}{2} & \frac{\Sigma-\Delta}{2}\\
\sqrt{\frac{1-\Sigma^{2}}{2}} & -i\sqrt{\frac{1-\Delta^{2}}{2}} & \frac{\Sigma-\Delta}{2} & -\frac{\Sigma+\Delta}{2}
\end{array}
\right)  ,\label{S_mat_symm_back_coup2}%
\end{equation}

\noindent where $\Sigma\equiv g+h$, $\Delta\equiv g-h$, and 

\begin{align}
g &  =\frac{\sqrt{1-t^{2}}}{1+\frac{\alpha^{2}}{\beta^{2}}t^{2}},\label{gWalpha}\\
h &  =\frac{\alpha}{\beta^{2}}\frac{t^{2}}{1+\frac{\alpha^{2}}{\beta^{2}}t^{2}}.\label{hWalpha}%
\end{align}

\noindent In the derivation of Eq.\,(\ref{S_mat_symm_back_coup2}), we assume, without loss of generality, that $\varphi_s=\pi/2$. 

When $\alpha=\beta=1/\sqrt{2}$, $\varphi=-\pi/2$, and the JPCs are biased at the $50:50$ beam splitter working point, the MPIJIS scattering matrix becomes

\begin{equation}
\left[  S\right]  =\left(
\begin{array}
[c]{cccc}%
0 & 0 & -\frac{1}{\sqrt{2}} & -\frac{1}{\sqrt{2}}\\
\frac{i2\sqrt{2}}{3} & 0 & -\frac{i}{3\sqrt{2}} & \frac{i}{3\sqrt{2}}\\
-\frac{1}{3\sqrt{2}} & -\frac{i}{\sqrt{2}} & -\frac{\sqrt{2}}{3} & \frac{\sqrt{2}}{3} \\
\frac{1}{3\sqrt{2}} & -\frac{i}{\sqrt{2}} & \frac{\sqrt{2}}{3} & -\frac{\sqrt{2}}{3}
\end{array}
\right) \cdot\label{S_mat_t_gh}
\end{equation}

A graphical representation of the scattering parameters of Eq.\,(\ref{S_mat_t_gh}) is displayed in Fig.\,\ref{SignalFlow}b. This result shows that when the MPIJIS is operated at this working point, it functions as an isolator with vanishing reflections $|S_{11}|=|S_{22}|=0$, almost unity transmission in the forward direction $|S_{21}|=2\sqrt{2}/3\cong0.943$ (which corresponds to an insertion loss of about $0.5$ dB in the signal power), and total isolation $|S_{12}|=0$. Furthermore, it shows that the cold loads on ports $3$ and $4$ play a similar role to internal ports of standard magnetic isolators. They dissipate the energy of back-propagating signals $|S_{32}|=|S_{42}|=1/\sqrt{2}$ and emit noise (e.g., vacuum noise) towards the input $|S_{13}|=|S_{14}|=1/\sqrt{2}$. 
  
Another interesting case occurs when $t=\beta$ and $r=\alpha$. In this case, we get $|S_{21}|=2\alpha/(1+\alpha^2)$. This result shows that $2\sqrt{2}/3$ is not a fundamental upper bound on $|S_{21}|$. In fact, for a highly uneven effective coupler with $\alpha\approx1$, $\beta\approx0$ and JPCs operated with little conversion $r\rightarrow{1}$, $t\rightarrow{0}$, we obtain $|S_{21}|\rightarrow1$. 

To generalize the on-resonance-derived scattering parameters listed in Eqs.\,(\ref{S21})-(\ref{S42}) for signals within the device bandwidth, we substitute \cite{JPCreview}

\begin{align}
t[\omega_1]=\dfrac{2\rho}{\chi_{a}^{-1}\chi_{b}^{-1}+\rho^2}, \label{t_param_vs_freq} 
\end{align}

\noindent where $\chi's$ are the bare response functions of modes \textit{a} and \textit{b} (whose inverses depend linearly on $f_1$ and $f_2$): 

\begin{align}
\chi_{a}^{-1}[\omega_{1}]=1-2i\dfrac{\omega_{1}-\omega_{a}}{\kappa_{a}}, \nonumber \\ 
\chi_{b}^{-1}[\omega_{2}]=1-2i\dfrac{\omega_{2}-\omega_{b}}{\kappa_{b}}. 
\label{Chi_params}%
\end{align}

Since the applied pump frequency satisfies the relations $\omega_{p}=\omega_{b}-\omega_{a}=\omega_{2}-\omega_{1}$, $\chi_{b}^{-1}$ of Eq.\,(\ref{Chi_params}) can be rewritten as  $\chi_{b}^{-1}[\omega_{1}]=1-2i(\omega_{1}-\omega_{a})/\kappa_{b}$. Obviously, this generalization holds under the assumption that the bandwidth of the $90^{\circ}$ hybrid is much larger than the dynamical bandwidths of the JPCs, which is generally the case because transmission-line-based hybrids exhibit bandwidths of a few hundreds of megahertz \cite{CPWhybrids}. 

In Fig.\,\ref{ThyFig}, we present several key theoretical results of the MPIJIS. Figure\,\ref{ThyFig}a shows the transmission parameters magnitude $|S_{21}|^2$ (blue) and $|S_{12}|^2$ (orange) of the MPIJIS, operated in the forward direction, as a function of normalized frequency, where $\omega_0=\omega_a$. As expected, $|S_{12}|^2$ exhibits a large dip in vicinity of the resonance while $|S_{21}|^2$ remains very close to unity. Figure\,\ref{ThyFig}b displays $|S_{21}|^2$ and $|S_{12}|^2$ of a forward-operated MPIJIS on resonance as a function of the JPC transmission parameter \textit{t}. In Fig.\,\ref{ThyFig}c, we show the dependence of a rather comprehensive set of the scattering parameters on the pump phase difference $\varphi$, namely, $|S_{11}|^2$ (red), $|S_{21}|^2$ (blue), $|S_{12}|^2$ (orange), $|S_{31}|^2$ (dashed yellow), $|S_{23}|^2$ (dashed magenta), $|S_{13}|^2$ (dashed black), and $|S_{32}|^2$ (dashed green). Note that some of the absent scattering parameters are equal in magnitude to ones that are present, i.e., $|S_{22}|^2=|S_{11}|^2$, $|S_{41}|^2=|S_{31}|^2$, $|S_{24}|^2=|S_{23}|^2$, $|S_{14}|^2=|S_{13}|^2$, $|S_{42}|^2=|S_{32}|^2$. Similar to the calculation of Fig.\,\ref{ThyFig}a,b the different scattering parameters in Fig.\,\ref{ThyFig}c are calculated using Eqs.\,(\ref{S21}), (\ref{S12}), (\ref{S11}), (\ref{S13}), (\ref{S23}), (\ref{S31}), (\ref{S32}) for the case of a symmetrical effective coupler. Also, in Figs. \ref{ThyFig}a,c, the JPCs are operated at the $50:50$ beam splitter point. As expected for $\varphi=-\pi/2+2\pi k$, where $k$ is an integer, $|S_{21}|^2$ (forward transmission) assumes a maximum, while $|S_{11}|^2$ (reflection), $|S_{12}|^2$ (backward transmission) assume a minimum. Similarly, $|S_{13}|^2$ (transmission from the load to the input) and $|S_{32}|^2$ (transmission from the output to the load) assume a maximum, while $|S_{31}|^2$ (transmission from the input to the load) and $|S_{23}|^2$ (transmission from the load to the output) assume a minimum. The responses are completely reversed for $\varphi=\pi/2+2\pi k$. Finally, Fig.\,\ref{ThyFig}d displays a representative set of the scattering parameters of the MPIJIS on resonance versus the transmission amplitude between modes \textit{b} of the JPCs, i.e., $\alpha$. In this calculation, the MPIJIS is operated in the forward direction and the transmission parameter of the JPCs $t$ is adjusted for each chosen $\alpha$ ($t\cong\beta$) to yield a fixed attenuation of $20$ dB in the backward direction $|S_{12}|^2$ (not shown). As seen in the figure, $|S_{21}|^2$ asymptotically approaches $1$ in the limit $\alpha\rightarrow{1}$. Moreover, while $|S_{13}|^2$ (transmission from the load to the input), represented by the red line, remains close to $-3$ dB, $|S_{23}|^2$ (transmission from the load to the output), represented by the magenta line, approaches $0$ in the limit $\alpha\rightarrow{1}$.  

\noindent \textbf{Added noise.} To calculate the added noise by the MPIJIS operated in the forward direction, we compare the signal-to-noise ratio at the output $S_{\rm{2}}/N_{\rm{2}}$ to the signal-to-noise ratio at the input $S_{\rm{1}}/N_{\rm{1}}$, where $S_{\rm{i}}$ and $N_{\rm{i}}$ represent the number of signal and noise-equivalent photons per mode per unit time per unit bandwidth at port i, respectively \cite{JPCreview}. Using the full scattering matrix of the MPIJIS (Eq.\,(\ref{S_mat_MPIJIS})), we write $S_{\rm{2}}=|S_{21}|^2S_{\rm{1}}$, $N_{\rm{2}}=|S_{21}|^2N_{\rm{1}}+|S_{23}|^2N_{\rm{3}}+|S_{24}|^2N_{\rm{4}}=N_{\rm{1}}$, where the last equality holds because the scattering matrix is unitary and $S_{22}=0$. We also assume here that the dominant noise entering the system is vacuum noise. Using these relations, we obtain for the noise factor $\rm{NF}=(S_{\rm{2}}/N_{\rm{2}})/(S_{\rm{1}}/N_{\rm{1}})=|S_{21}|^2$. Alternatively, $\rm{NF}$ can be expressed in terms of the number of noise-equivalent photons added by the MPIJIS to the input $n_{\rm{add}}$, where $S_{\rm{2}}=|S_{21}|^2S_{\rm{1}}$ and $N_{\rm{2}}=|S_{21}|^2(N_{\rm{1}}+n_{\rm{add}})$. In this representation, $\rm{NF}=N_{\rm{1}}/(N_{\rm{1}}+n_{\rm{add}})$. Solving for $n_{\rm{add}}$, gives $n_{\rm{add}}=(1-|S_{21}|^2)/2|S_{21}|^2$, where we substituted the vacuum-noise contribution $N_{\rm{1}}=1/2$. Based on this result, we estimate the added noise by the MPIJIS in the qubit measurement of Fig.\,\ref{CohFidSparam}b to be $n_{\rm{add}}\backsimeq1/2$, obtained at $|S_{21}|^2=1/2$ (corresponding to about $3$~dB dip on resonance observed in Fig.\,\ref{CohFidSparam}b (i)).
 
\noindent \textbf{Characterization measurements of the MPIJIS.} In Figure\,\ref{SparamIsoMeas}, we exhibit scattering parameters measurement of the MPIJIS versus frequency taken while operated in the forward direction $\varphi=-\pi/2$ (Figs. \ref{SparamIsoMeas}a, \ref{SparamIsoMeas}c) and the backward direction $\varphi=\pi/2$ (Figs. \ref{SparamIsoMeas}b, \ref{SparamIsoMeas}d). In this measurement, taken in a separate cooldown, the MPIJIS is characterized directly without the qubit setup. The solid (dashed) curves represent measurements taken with the MPIJIS \textit{on} (\textit{off}). The blue, orange, red, and magenta curves represent $|S_{21}|^2$, $|S_{12}|^2$, $|S_{11}|^2$, and $|S_{22}|^2$, respectively. When the MPIJIS is \textit{off}, both transmission parameters are close to unity ($\approx 0$ dB) and the reflection parameters are suppressed (below $-10$ dB), which is consistent with the MPIJIS being transparent for transmitted signals between ports $1$ and $2$ in the \textit{off} state. When the MPIJIS is operated in the forward direction (Fig.\,\ref{SparamIsoMeas}a), $|S_{21}|^2$ remains relatively close to unity $\sim-1.8$ dB, while $|S_{12}|^2$ exhibits a dip of $-20.3$ dB on resonance and a dynamical bandwidth of $11$ MHz. When the MPIJIS is operated in the backward direction, the roles played by the transmission parameters $S_{21}$ and $S_{12}$ are reversed as seen in Fig.\,\ref{SparamIsoMeas}b. Moreover, the reflection parameters $|S_{11}|^2$, and $|S_{22}|^2$ are suppressed further on resonance, as expected, when the MPIJIS is \textit{on} as seen in Figs. \ref{SparamIsoMeas}c and \ref{SparamIsoMeas}d.

Furthermore, we measure the maximum input power which the MPIJIS can handle on resonance at a given isolation (e.g., $20$ dB), above which it saturates. The measurement result taken for the MPIJIS operated in the backward direction (same working point as Fig.\,\ref{SparamIsoMeas}b) is shown in Fig.\,\ref{DR}. The plot depicts the transmission parameters $|S_{21}|^2$ (blue) and $|S_{12}|^2$ (orange) as a function of input power. As seen in the figure, $|S_{12}|^2$ is almost flat around $-1.5$ dB up to $-100$ dBm, whereas $|S_{21}|^2$ is almost flat around $-22$ dB in the range between $-130$ to $-110$ dBm but it gradually degrades beyond $-110$ dBm. One figure of merit which we apply here to quantify the saturation power of the isolator is the 1 dB compression point (denoted as $P_{-1\rm{dB}}$), which is commonly used to characterize amplifiers. In the amplifier case, the $1$ dB compression point usually corresponds to the input power for which the low-input-power-gain of the amplifier degrades by $-1$ dB. Analogously, the $1$ dB compression point in our Josephson isolator case corresponds to the input power for which the isolation degrades by $+1$ dB (denoted as $P_{+1\rm{dB}}$). In our device, we find $P_{+1\rm{dB}}=-108$ dBm, which is indicated by a red circle in Fig.\,\ref{DR}. It is worth noting that this figure is significantly larger than $P_{-1\rm{dB}}$ of microstip JPCs, when operated in the amplification mode, which is on the order of $-130$ dBm \cite{JPCreview,hybridLessJPC}. This is because in the isolator case the JPCs are operated in the frequency conversion mode (without photon gain) \cite{Conv}.

Finally, we present in Fig.\,\ref{Spectrum} a broadband spectrum measured at the output of one of the MPIJIS ports. The black and blue curves are data taken using a spectrum analyzer while the MPIJIS is in the \textit{off} and \textit{on} states, respectively. In the \textit{on} state, the MPIJIS is operated at the same working point as in Fig.\,\ref{SparamIsoMeas}b. As seen in the figure, the MPIJIS does not generate any spurious harmonics or noise when it is \textit{on}, however, it does allow some pump power injected through the pump ports to leak out through the device ports $1$ and $2$ as indicated by the peak observed at $f_p=2.8$ GHz. This observation could potentially explain the slightly increased dephasing experienced by the qubit when the MPIJIS is \textit{on} (Fig.\,\ref{CohFidSparam}b) versus \textit{off} (Fig.\,\ref{CohFidSparam}a).  

\noindent \textbf{Qubit measurement parameters.} The frequencies of the pumps applied to the JPC (amplifier) and MPIJIS, in the measurements of Fig.\,\ref{CohFidSparam}, are $16.78$ GHz and $2.771$ GHz, respectively. All qubit data exhibited in Fig.\,\ref{CohFidSparam} is taken with a readout pulse duration of $t_{\rm{r}}=200$ ns, integration time of $t_{\rm{int}}=150$ ns, and an average photon number in the readout resonator of $\bar{n}=6.4$. The qubit data is averaged over $2000$ iterations. Using the expression $\rm{tan}(\theta/2)=\chi/\kappa$, we extract a qubit-state-dependent phase shift of the readout signal of $\theta=46^{\circ}$. Furthermore, by substituting the histogram peak location $\bar{\rm{I}}/\sigma=2.2$ for configuration \textbf{c} in Fig.\,\ref{CohFidSparam} (in which the JPC is \textit{on}) in the relation $\bar{\rm{I}}/\sigma=\sqrt{2\bar{n} \eta \kappa t_{\rm{int}}} \rm{sin}(\theta/2)$ \cite{QubitJPC}, we get an approximate value for the measurement efficiency of our readout chain $\eta\backsimeq0.32$.   

\noindent \textbf{Comparison to conventional high-fidelity qubit setup.} In Fig.\,\ref{TypHFmeas}a we exhibit a block-circuit diagram of a conventional rapid, high-fidelity qubit readout setup, which employs magnetic-based circulators and isolators. In Fig.\,\ref{TypHFmeas}b-d we exhibit qubit measurements corresponding to qubit readout fidelity, $T_1$, and $T_{\rm{2E}}$, respectively, taken in a separate cooldown with the conventional setup displayed in Fig.\,\ref{TypHFmeas}a. All three measurements shown in Figs.\,\ref{TypHFmeas}b-d are taken with the JPC turned off using the same qubit-resonator-Purcell filter system and measurement parameters of Fig.\,\ref{CohFidSparam}. As seen in Figs.\,\ref{TypHFmeas}b-d, the measured readout fidelity $F=0.75$, qubit lifetime $T_1=49$ $\mu$s, and coherence time $T_{\rm{2E}}=25$ $\mu$s are comparable with those measured for configuration \textbf{a} of Fig.\, \ref{CohFidSparam}, which uses the superconducting directional coupler and MPIJIS instead of the qubit-circulator and intermediate isolator shown in Fig.\,\ref{TypHFmeas}a. Note that although $T_1=49$ $\mu$s measured with the conventional setup of Fig.\,\ref{TypHFmeas}a is slightly higher than $T_1=39$ $\mu$s measured with the new setup of Fig.\,\ref{MeasSetup}, they fall within the variation range observed for different cooldowns. It is worth noting that turning on the JPC in this conventional setup, which has three isolating stages between the JPC and the qubit-resonator, does not degrade $T_1$ or $T_{\rm{2E}}$ shown in Figs.\,\ref{TypHFmeas}c,d. It only enhances the measurement fidelity shown in Fig.\,\ref{TypHFmeas}b to about $F=0.99$.  

\noindent \textbf{Scattering parameters of the directional coupler.} The superconducting, on-chip directional coupler is realized using coupled coplanar waveguides. The characterization results of the on-chip, wideband, superconducting directional coupler are shown in Fig.\,\ref{DircCoupRes}. A circuit symbol of the directional coupler is shown in the upper-left corner of the figure, which defines the four ports of the device. The characterization measurement is done in a separate cooldown without the qubit setup, where each port of the directional couple is connected to a three-port, cryogenic circulator with its own input and output lines. A schematic image of the directional coupler chip mounted into a $50$ $\Omega$-matched pogo-package is shown in the upper-right corner. As seen in the image the pogos (pins) inside the package connect the four ports of the directional coupler to designated copper traces in a multi-layer PCB. The four copper traces, carrying incoming and outgoing signals, connect to surface-mount SMA connectors at the periphery of the PCB (not shown), which in our case define the directional coupler ports. In Figs. \ref{DircCoupRes}a-d, we exhibit calibrated measurements of the $16$ scattering parameters of the directional coupler in the $4-8$ GHz frequency range (limited by the bandwidth of cryogenic, three-port circulators employed in the measurement). Figure\,\ref{DircCoupRes}a exhibits the measured magnitude of the transmission parameters $|S_{21}|^2$ (blue), $|S_{12}|^2$ (orange), $|S_{34}|^2$ (magenta), $|S_{43}|^2$ (red). The data shows that the transmission through the device is close to $0$ dB in a wide bandwidth except for a narrow window around $6.5$ GHz, where the transmission between ports $3$ and $4$ is around $-2$ dB. It is worth noting that this measurement includes the insertion loss of the multi-layer PCB, whose dielectric material is FR4 ($\epsilon_d=3.65$). Figure\,\ref{DircCoupRes}b displays the measured coupling parameters $|S_{31}|^2$ (blue), $|S_{13}|^2$ (orange), $|S_{24}|^2$ (magenta), $|S_{42}|^2$ (red). The data shows that the coupling of the device is approximately flat across the bandwidth with an average value of about $-19$ dB. Figure\,\ref{DircCoupRes}c plots the measured reflection parameters $|S_{11}|^2$ (blue), $|S_{22}|^2$ (orange), $|S_{44}|^2$ (magenta), $|S_{33}|^2$ (red). The data shows that the reflection off the device ports varies across the measurement bandwidth in the $-10$ dB to $-20$ dB range. Finally, Fig.\,\ref{DircCoupRes}d depicts the isolation parameters of the device $|S_{41}|^2$ (blue), $|S_{14}|^2$ (orange), $|S_{23}|^2$ (magenta), $|S_{32}|^2$ (red). As seen in the figure, the isolation varies between $-25$ to $-40$ dB in the measurement bandwidth but averages around $-30$ dB.

\noindent \textbf{Directional coupler package.} 
To preserve the wideband characteristics of the designed superconducting directional coupler when coupling its pads to $50$ $\Omega$ PCB transmission lines, we utilize a pogo-package technology detailed in Ref. \cite{QST_Bronn2018}, which provides good vertical $50$ $\Omega$ impedance-matched transitions compared to traditional wirebond technology. A blow-up of the pogo-package used is shown in Fig.\,\ref{Pogo}. It consists of, from bottom to top, a pedestal, an extrusion, a ground board, and an interposer. A cutout is included in the pedestal to push any chip modes to high frequency, and the chip is aligned to the two bosses on the extrusion. The ground board is made to press down on the edges for thermalization and has plating on the inside surfaces of the cutout to avoid exposed dielectric. The tolerances are adjusted so the chip protrudes slightly above the surface of the extrusion to establish positive contact with the ground board. Wirebonds are then added from the ground board to the chip ground plane and across the traces to suppress parasitic modes. The interposer is then clamped on top and spring-loaded pins with $50$ $\Omega$ dielectrics are used for electrical connections, as described in \cite{QST_Bronn2018}. 

\noindent \textbf{The Purcell filter.} The Purcell filter is realized using five sections of stepped impedance coplanar waveguide (CPW) transmission lines \cite{Bronn2015b}. The CPW sections have alternating characteristic impedances and lengths, i.e., $Z_\mathrm{lo} = 25$~$\Omega$ of length $\ell_\mathrm{lo} = 8.5$~mm, and $Z_\mathrm{hi} = 120$~$\Omega$ of length $\ell_\mathrm{hi} = 6.5$~mm. The Purcell filter starts and ends with a low-impedance section. The device functions as a bandpass filter for readout frequencies spanning about $1$ GHz of bandwidth around $6.5$ GHz. Moreover, it suppresses signals in the frequency range $2$-$6$ GHz. More specifically, it yields an attenuation of about $20$ dB near the qubit frequency ($\sim 5.2$~GHz) \cite{Bronn2015b}. The Purcell filter is back-mounted into the same PCB as the qubit-resonator chip using a copper bottom cover. The qubit-resonator chip and the Purcell filter are coupled through a $\sim10$ mm long 50~$\Omega$ stripline transmission line within the PCB.

\noindent \textbf{Fabrication process.} The qubit-resonator circuit is fabricated on high-resistivity silicon by mixed optical and e-beam lithography. Large features of the superconducting quantum circuit are transferred into a thin film of niobium by optical lithography and reactive ion etch. Subsequently, the Al-Al$_2$O$_3$-Al Josephson junctions are defined by electron beam lithography and deposited by shadow mask evaporation in a Dolan-bridge process.

The directional coupler chip is fabricated with a single-layer optical lithography process using sputter-deposited Nb metal on a $3$-inch, high-resistivity, float zone silicon substrate. Native oxide is removed from the substrate \textit{in situ} with an Ar plasma immediately prior to deposition. The photoresist mask is then patterned on the surface of the Nb and the coplanar gaps are etched with SF$_6$ plasma using laser-reflection endpoint detection. The photoresist mask is stripped in hot solvents. Finally, a protective coat of resist is applied before the wafer is diced into dies.

The Purcell filter is implemented using a SF$_6$ subtractive dry etch of 200~nm thick niobium sputtered on a $4 \times 10$~mm$^2$ sapphire substrate.

\noindent \textbf{Acknowledgments} 
B.A. highly appreciates J. Rozen's help in wiring the dilution fridge. Work pertaining to the development of the Purcell filter was supported by IARPA under contract W911NF-10-1-0324, and to the development of the pogo-pin packaging by IARPA under contract W911NF-16-1-0114-FE. Contribution of the U.S. Government, not subject to copyright.

\noindent \textbf{Author Contributions} 
B.A. designed the isolator device, performed the experiment, and analyzed the data. N.B. designed the Purcell filter and the PCBs, O.J. designed the device package, coils and mounting brackets, S.O. designed and simulated the directional coupler, A.D.C. assembled the qubit measurement setup, M.B., V.P.A., R.E.L. fabricated the devices, X.W. performed preliminary characterization of the directional couplers, D.P.P. designed and provided the package for the directional coupler, J.M.C. supervised the project. B.A. wrote the paper with input and contributions from the other authors.

\end{document}